\documentclass[aps,prd,twocolumn]{revtex4-2}
\usepackage{comment}
\usepackage{amsmath}
\usepackage{ulem}
\usepackage{color}
\usepackage{graphicx}
\usepackage{epsfig}
\usepackage{subfig}
\usepackage{bm}
\usepackage{hyperref}
\usepackage{natbib}

\def\be {\begin{equation}}
\def\ee {\end{equation}}
\def\nn {\nonumber}
\def\bea {\begin{eqnarray}}
\def\eea {\end{eqnarray}}

\begin{document}
\title{Anisotropy of magnetized quark matter}

\author{Kangkan Goswami}
\author{Dushmanta Sahu}
\author{Jayanta Dey}
\author{Raghunath Sahoo}
\email{Corresponding Author: Raghunath.Sahoo@cern.ch}
\affiliation{Department of Physics, Indian Institute of Technology Indore, Simrol, Indore 453552, India}

\author{Reinhard Stock}
\affiliation{Institute of Nuclear Physics, University of Frankfurt, Germany}

\begin{abstract}
    Strong transient magnetic fields are generated in non-central relativistic heavy-ion collisions. These fields induce anisotropy within the strongly interacting medium that, in principle, can affect the thermodynamic properties of the medium. We use the Polyakov loop extended Nambu Jona-Lasinio model to study the quark matter subjected to an external magnetic field at vanishing baryon chemical potential ($\mu_{B}$). We have estimated the degree of anisotropy in the speed of sound and isothermal compressibility within the magnetized quark matter as a function of temperature ($T$) and magnetic field ($eB$). This study helps us to understand the extent of directionality generated in the initial stages of non-central collisions while giving us useful information about the system.
\end{abstract}
\date{\today}
\maketitle
\section{Introduction}

One of the primary goals of relativistic heavy-ion collisions is to study the deconfined state of strongly interacting quarks and gluons in local thermal equilibrium, known as the quark-gluon plasma (QGP). 
In the non-central heavy-ion collision, the charged spectators move past the fireball at a relativistic speed. According to the Biot-Savart law, these moving charge particles create a large transient electromagnetic field of the order of $10^{18}$ G at the Large Hadron Collider (LHC)~\cite{Deng:2012pc, Skokov:2009qp}. 
Direct experimental evidence for the strength of the magnetic field ($B$) is yet to be discovered. However, recent measurements of the directed flow of $D^0$ and $\bar{D}^0$ at the Relativistic Heavy Ion Collider (RHIC)~\cite{STAR:2019clv} and Large Hadron Collider~\cite{ALICE:2019sgg}, indicate the creation of a strong magnetic field during the collision.

The first principle non-perturbative theory of strong interaction, lattice quantum chromodynamics (lQCD), found many interesting phenomena in QGP in the presence of magnetic field such as chiral magnetic effect~\cite{Fukushima:2008xe}, magnetic and inverse magnetic catalysis~\cite{Bali:2011qj}. However, experimental verification of these phenomena are yet to be observed. Under an external magnetic field, the energy levels of the charged particles get quantized following the Landau quantization, which creates momentum anisotropy affecting various thermodynamical~\cite{Tawfik:2016lih, Bali:2014kia, Ferreira:2013oda} and dissipative quantities~\cite{PhysRevD.102.114015, PhysRevD.102.016016, PhysRevC.106.044914}. For instance, the thermodynamical pressure becomes anisotropic with a longitudinal component along the magnetic field and a transverse component in the transverse plane of the field. This leads to anisotropy in many other thermodynamical quantities, such as speed of sound ($c_s$), and isothermal compressibility ($\kappa_{\rm {T}}$). Very recently, the anisotropy in the speed of sound in a magnetized hybrid neutron star is explored in Ref.~\cite{Ferrer:2022afu}. The authors consider a magnetized hybrid neutron star with three density phases, employ three different models to study the speed of sound and explore the anisotropy created due to a magnetic field at zero temperature. The MIT Bag model is considered for high-density regimes where quark matter could possibly exist. Ref.~\cite{Yang:2021rdo} also explored the anisotropy in equation of state for magnetized baryon matter at finite temperature and chemical potential using a two-flavor Nambu Jona-Lasinio (NJL) model. The study aims to locate the critical end-point with the help of isothermal compressibility and quark number susceptibility. Such studies in low-temperature and high-$\mu_{\rm{B}}$ regimes motivate us to understand and quantify the anisotropy generated in peripheral heavy-ion collision at the LHC energies. In the strong field limit, Landau quantization leads to a dimensional reduction in the phase space ($3 \rightarrow 1$ dimension)  at the lowest Landau level (LLL). Moreover, the magnetic field influences the QCD phase diagram. Different effective QCD models are used to study the phase diagram in the $B-T$ plane, such as linear sigma model~\cite{PhysRevD.82.105016}, NJL model, and its extended version Polyakov loop extended NJL (PNJL) model~\cite{PhysRevD.82.054027, Gatto:2010pt}. Initial prediction of the PNJL model showed that the transition temperature ($T_c$) and its strength increase with $eB$, leading to a first-order phase transition. However, lQCD calculation~\cite{Bali:2011qj} found an opposite trend, which means $T_c$ decreases with $eB$. The same was also reproduced in model calculations such as NJL and PNJL models~\cite{Ferreira:2014kpa, Tavares:2021fik} with a magnetic field-dependent coupling constant. Magnetic field-dependent coupling constant was introduced in these models to allow them to explain the phenomena of the inverse magnetic catalysis, as observed in lQCD.

Many indirect probes and observables are suggested to study the microscopic and bulk features of QGP under the effect of the magnetic field. Theoretically, one can study the change in the thermodynamic observables to understand the changes in the deconfined medium in the presence of an external magnetic field. The behavior of certain thermodynamic observables such as $\kappa_{\rm T}$ and $c_{\rm s}^{2}$ provide useful information about the nature of the phase transition of the system. $\kappa_{\rm T}$ represents the rate of change in the volume of the system concerning pressure at a constant temperature. Precisely, $\kappa_{\rm T}$ measures the extent to which the density of quarks and gluons changes in response to changes in external pressure, which is an essential factor in determining the equation of state of the medium~\cite{STOKIC2009192, MEKJIAN2005239, MUKHERJEE20181}. Moreover, it can tell us about the degree of deviation of a system from a perfect fluid. $\kappa_{\rm{T}}$ is expected to show a sudden jump near the critical end-point (CEP), where the smooth crossover in the QCD phase diagram meets the first-order phase transition. Thus, it is an interesting observable to explore in the QCD phase diagram~\cite{Karsch2002}. In literature, $\kappa_{\rm{T}}$ has been studied as a function of temperature and charged particle multiplicity~\cite{Jain:2021qro, Sahu:2020nbu, Khuntia:2018non, Pal:2021qav, Mukherjee:2017elm, Sahu:2020swd}. In Ref.~\cite{Sahu:2020nbu}, the high-temperature QCD matter has been found to be the closest to a perfect fluid.

On the other hand, the speed of sound reflects the propagation of small perturbations produced in the system in its local rest frame. Its dependence on the environment, i.e. temperature, density, and baryon chemical potential, means that it is an ideal probe to explore the evolution of the fireball~\cite{Albright:2015fpa}. The studies in Refs.~\cite{ Campanini:2011bj, Gardim:2019xjs, Sahu:2020swd, Biswas:2019wtp} reveal that exploring $c_{\rm s}^{2}$ as a function of charged particle multiplicity can be utilized to study the dynamics of heavy-ion collisions. Moreover, it shows minima near the phase transition. This also makes $c_{\rm s}^{2}$ a suitable observable to study the QCD phase space. In literature, many phenomenological models are employed to compute the speed of sound such as the quasiparticle model \cite{Mykhaylova:2020pfk}, dynamic quasiparticle model \cite{Soloveva:2021quj}, color string percolation model \cite{Sahu:2020mzo, Sahoo:2017umy}, hadron resonance gas model \cite{Venugopalan:1992hy, Bluhm:2013yga}, NJL model \cite{Deb:2016myz, Marty:2013ita},  and PNJL model \cite{Motta:2020cbr, Ghosh:2006qh, He:2022kbc}. It has also been estimated using the lQCD calculations \cite{Aoki:2006we, HotQCD:2014kol, Borsanyi:2013bia, Borsanyi:2020fev}. 

The lattice QCD model is a first principle theory of strongly interacting matter. Despite its many successes, it fails to explain the high-baryon-rich environment due to its fermion sign problem~\cite{Karsch2002}. Hence, exploring the CEP within the lQCD formalism is not feasible. Thus, the NJL and PNJL models are good alternatives in this regard. The PNJL model is an extended version of the NJL model, initially developed for nuclear matter. Due to its respect for the global symmetries of QCD, particularly the chiral symmetry, this model has been widely utilized to explore some of the non-perturbative features of the QCD vacuum. The NJL model is not renormalizable due to the point-like interaction between the quarks \cite{Nambu:1961tp, Nambu:1961fr}. A suitable regularisation method must be established to deal with the divergent integrals. The three-momentum cutoff scheme is widely accepted among many others in the literature~\cite{Klevansky:1992qe}. The model's parameters must be fixed to replicate well-known phenomenological quantities, such as the pion-decay constant and quark condensate density~\cite{Klevansky:1992qe}. The NJL model is based on the idea that the chiral symmetry of quantum chromodynamics (QCD) is spontaneously broken at low temperatures and densities, which leads to the appearance of a non-zero quark condensate. The NJL model incorporates this symmetry breaking by introducing a four-fermion interaction between quarks. The interaction is attractive in the scalar channel and repulsive in the pseudoscalar channel, which leads to the formation of a chiral condensate and the generation of quark masses~\cite{Meisinger:2001fi}.

Albeit being a successful model, one drawback of the NJL model is that it doesn't incorporate the gluon dynamics, and as a result, quark confinement is absent in this model. As suggested in~\cite{Fukushima:2003fw}, the Polyakov loop characterizes the effects of confinement, which prevents quarks from existing as free particles. The trace of the Polyakov loop ($\rm \Phi$) is the order parameter for the confinement transition in a pure gauge theory~\cite{Polyakov:1978vu, Susskind:1979up, Svetitsky:1982gs, Svetitsky:1985ye}. It can be understood as $\Phi$ is related to $e^{-F/T}$, where $F$ is the free energy of the static quark. The free energy of a confined single quark is infinite, which leads to $\Phi = 0$, and in the deconfined phase, free energy is finite, which makes $\Phi \neq 0$~\cite{Fukushima:2011jc}. In the PNJL model, the Polyakov loop is included as a background temporal gluon field that interacts with the quarks. The PNJL model has been successful in reproducing lattice results qualitatively~\cite{Ratti:2005jh, Ghosh:2006qh, Mukherjee:2006hq, Ratti:2007jf, Albright:2015fpa}. The PNJL model has also been used to study a wide range of phenomena in QCD matter, including thermodynamic and transport properties \cite{Bhattacharyya:2010wp, Bhattacharyya:2016jsn, Bhattacharyya:2017gwt, Bhattacharyya:2019qhm, Islam:2019tlo, Soloveva:2021quj}, probing the critical point in the $(T-\mu)$ plane~\cite{Roessner:2006xn, Fukushima:2008wg, Carignano:2010ac, Hell:2008cc}, and exploring the QCD medium under the effect of an external magnetic field~\cite{Ferreira:2013oda, Ferreira:2013tba, Ferreira:2014kpa, Chaudhuri:2022oru, Wang:2020mww}.

In this work, we investigate the behavior of the speed of sound and isothermal compressibility and explore the degree of anisotropy generated in baryon-free quark matter in the presence of an external magnetic field. This paper is organized as follows. In section \ref{formulation B0} and \ref{formulation Bneq0}, we have discussed the PNJL model formulation with and without an external magnetic field, respectively. Section \ref{sec result} comprises the results of this work. Finally, section \ref{summary} summarizes this work.

\section{Formulation}
\subsection{Quark matter in zero magnetic field}
\label{formulation B0}
The Polyakov loop extended Nambu-Jona Lasinio (PNJL) model was introduced as an improvement to the Nambu-Jona Lasinio (NJL) model \cite{ Meisinger:1995ih, Meisinger:1995kp, Fukushima:2003fw, Ratti:2005jh}. The NJL model effectively takes care of the chiral symmetry breaking. However, due to the lack of gluonic interaction in the model, it fails to explain the deconfinement dynamics of quarks. PNJL model takes care of this problem by adding a Polyakov loop effective potential to the NJL Lagrangian. This model simplifies the interaction between quarks and gluons by considering chiral point couplings between quarks and a temporal background gauge field. Extensive studies have been done using the PNJL model with 2 and 2+1 flavors~\cite{Ratti:2005jh, Roessner:2006xn, Fukushima:2008wg, Ghosh:2006qh, Hansen:2006ee, Gatto:2010pt, Fukushima:2010fe, Fu:2007xc}.
The PNJL lagrangian for 2+1 flavors is given by \cite{Fu:2007xc, Ciminale:2007sr, Fukushima:2008wg},
\begin{widetext}
\begin{equation}
\begin{split}
    {\mathcal L_{PNJL}} = \bar q(i \gamma^\mu D_\mu-\hat m)q + G\sum_{a=0}^{8} \left[ (\bar q\lambda^a q)^{2} + (\bar q i \gamma_5\lambda^a q)^{2} \right] - K\{{\rm det}[\bar q(1+\gamma_5)q] +\mbox{det}[\bar q(1-\gamma_5)q]\} - \mathcal{U}( \Phi, \overline{\Phi}, T)
\end{split}
\end{equation}
\end{widetext}
where $q$ is the three-flavor quark field, $q = (u,d,s)$, $\lambda^{a}$ are the Gell-Mann matrices in flavor space, and $\hat m$ is the current quark mass, $\hat m$ = diag $(m_{u}, m_{d}, m_{s})$. The covariant derivative is defined as,
\begin{equation*}
    D^{\mu}=\partial^{\mu}-i A^{\mu},
\end{equation*}
where $A^{\mu}=\delta^{\mu}_{0}A^0$ and in the Euclidian notation, we can write $A^{0} = -\textit{i} A_{4}$. The four-point interaction and six-point interaction of the quark field are incorporated with an effective coupling strength $G$ and $K$, respectively. $\mathcal{U}( \Phi, \overline{\Phi}, T)$ is the effective Polyakov loop potential.
The parameters in the Lagrangian are the cut-off parameter $\rm \lambda$, the coupling constants $\rm G$ and $\rm K$, and the current quark masses $\rm m_{u}$ and $\rm m_{s}$, we use the value set in Ref.~\cite{Rehberg:1995kh} as $\lambda = 602.3$ MeV, $G\lambda^{2} = 1.835$, $K\lambda^{5} = 12.36$, $m_{u} = m_{d} = 5.5$ MeV and $m_{s} = 140.7$ MeV. These parameters are obtained by fixing the values of $m_{\pi} = 135$ MeV, $m_{K} = 497.7$ MeV, $m_{\eta'} = 957.8$ MeV and $f_{\pi} = 92.4$ MeV.

This model takes the analogy from the BCS theory of superconductor where pairing between electrons in a spin-singlet state leads to a condensation which introduces a gap in the energy spectra. Similarly, the condensation $\langle \bar{q}_{R}q_{L}\rangle$ (or $\langle \bar{q}_{L}q_{R}\rangle$) arises due to the pairing between the quark-antiquark of the same chirality. Thus, $\langle \bar{q}{q} \rangle$ = $\langle \bar{q}_{R}q_{L} + \bar{q}_{L}q_{R}\rangle \sim \sigma$ can be taken as the order parameter and in the mean field approximation gives rise to a dynamical mass $M \sim \langle \bar{q}{q} \rangle$~\cite{Hatsuda:1994pi}. Thus, the gap equation for the PNJL model can be expressed as~\cite{Fu:2007xc},
\begin{equation}
\label{gap}
    M_{i} = m_{i} - 4G\sigma_{i} + 2K\sigma_{j}\sigma_{k}
\end{equation}
where $i,j,k = u,d,s$ in cyclic order, $M_{i}$ is the constituent quark mass, $m_{i}$ is the bare quark mass, and $\sigma_{i,j,k}$ is the quark condensate for different flavors. The third term exists as a result of introducing six-fermion interactions and in turn, generating a flavor mixing in the quark mass.


The thermodynamic potential is given as~\cite{Fu:2007xc, Torres-Rincon:2017zbr},
\begin{widetext}
\begin{equation} 
\begin{split}
\label{eq1}
    \Omega = \mathcal{U}( \Phi, \overline{\Phi}, T)  + 2{ G}\sum_{{f = u,d,s}}\sigma_{f}^2  -  4{ K}\sigma_u \sigma_d \sigma_s  -  2{N_{c}}\sum_{{f = u,d,s}} \int_{\Lambda} \frac{d^{3}p}{(2\pi)^{3}}\sqrt{p^{2} + M^{2}_{f}}  \\
    - 2T\sum_{{f =u,d,s}}\int \frac{d^{3}p}{(2\pi)^{3}}\rm{ln} \left[1 + 3\Phi e^{-\beta (E-\mu)} + 3\bar{\Phi} e^{-2\beta (E-\mu)} + e^{-3\beta (E-\mu)} \right] \\
    - 2T\sum_{{f =u,d,s}}\int \frac{d^{3}p}{(2\pi)^{3}}\rm{ln} \left[1 + 3\bar{\Phi} e^{-\beta (E+\mu)} + 3\Phi e^{-2\beta (E+\mu)} + e^{-3\beta (E+\mu)} \right]~,  
\end{split}
\end{equation}
\end{widetext}
where $\beta$ = 1/T, $E = \sqrt{p^{2} + M^{2}}$ is the energy of the quark and $\mathcal{U}( \Phi, \overline{\Phi}, T)$ is the effective Polyakov loop potential for the $\rm \Phi$ and $\rm \bar{\Phi}$ fields. Different versions of this potential exist in the literature \cite{Roessner:2006xn, Sasaki:2006ww, Ghosh:2007wy}. For our study, we choose the following form of the potential, which mathematically limits the value of $\Phi$ to be unity as $T~\rightarrow \infty$~\cite{Roessner:2006xn}, 
\begin{equation}
\begin{split}
\label{eq2}
    \frac{\mathcal{U}( \Phi, \overline{\Phi}, T)}{T^{4}} = - \frac{a(T)}{2}\Phi\overline{\Phi} +  b(T)\rm{ln}[ 1 - \\ 6\Phi\overline{\Phi} + 4 (\Phi^{3} + \overline{\Phi}^{3}) - 3(\Phi\overline{\Phi})^{2}]~, 
\end{split}
\end{equation}
where $a(T)$ and $b(T)$ reads,
\begin{eqnarray}
   a(T) = a_{0} + a_{1} \left( \frac{T_{0}}{T}\right) + a_{2}\left( \frac{T_{0}}{T}\right)^{2}~, 
   \nn\\ 
   b(T) = b_{3}\left( \frac{T_{0}}{T}\right)^{3}~,
\end{eqnarray}
where the parameters $a_{0}$, $a_{1}$, $a_{2}$ and $b_{3}$ are fixed by performing a simultaneous fit of the thermodynamic observables obtained from lattice QCD calculation and given as \cite{Roessner:2006xn},
\begin{equation}
\begin{split}
    a_{0} = 3.51,~a_{1} = -2.47 \\
    a_{2} = 15.2,~b_{3} = -1.75
\end{split}
\end{equation}
The parameter $ T_{0}$ is the critical temperature for the deconfinement phase transition in a pure-gauge system. In a pure gauge approach, $T_{0}$ is fixed to be 270 MeV. One can find different values of $T_{0}$ in literature. In Ref.~\cite{Schaefer:2007pw}, a $N_{f}$ dependent form of $T_{0}$ is taken. In Ref.~\cite{Zhang:2018ouu}, the authors studied the effect of different $T_{0}$ values on the thermodynamic observables.
However, following Ref.~\cite{Ratti:2005jh}, for quantitative comparison of results with lQCD data, we choose $T_{0}$ = 190 MeV.

In order to estimate $\Phi$, $\bar{\Phi}$ and $\sigma_{f}$, one needs to minimize $\Omega$ with respect to the above-mentioned quantities~\cite{Fu:2007xc, Costa:2008dp},
\begin{equation*}
    \frac{\partial \Omega}{\partial \sigma_f} = 0,   ~\frac{\partial \Omega}{\partial \Phi} = 0,   ~\frac{\partial \Omega}{\partial \bar{\Phi}} = 0
\end{equation*}
Solving $\frac{\partial \Omega}{\partial \sigma_f} = 0$, we get the condensates as~\cite{Costa:2008dp},
\begin{equation}
    \sigma_{f} = -2N_{c} \int_{\Lambda} \frac{d^{3}p}{(2\pi)^{3}}\frac{M_{f}}{E_{f}}(1- f_{\Phi}^{+} - f_{\Phi}^{-})
\end{equation}
where $f_{\Phi}^{+}$ and $f_{\Phi}^{-}$ are the quark and antiquark distribution functions respectively~\cite{Hansen:2006ee, Torres-Rincon:2017zbr},
\begin{equation*}
f_{\Phi}^{+} = \frac{(\Phi + 2\overline{\Phi}e^{-\beta(E_{f}-\mu_{f})})e^{-\beta(E_{f}-\mu_{f})} + e^{-\beta(E_{f}-\mu_{f})}} {1 + 3 \left(\Phi + \overline{\Phi} e^{-\beta(E_{f}-\mu_{f})}  \right) e^{-\beta(E_{f}-\mu_{f})} +  e^{-3\beta(E_{f}-\mu_{f})}},
\end{equation*}
\begin{equation*}
f_{\Phi}^{-} = \frac{(\overline{\Phi} + 2\Phi e^{-\beta(E_{f}+\mu_{f})})e^{-\beta(E_{f}+\mu_{f})} + e^{-\beta(E_{f}+\mu_{f})}} {1 + 3 \left( \overline{\Phi} + \Phi e^{-\beta(E_{f}+\mu_{f})} \right) e^{-\beta(E_{f}+\mu_{f})} +  e^{-3\beta(E_{f}+\mu_{f})}}.
\end{equation*}

Finally, pressure can be estimated as $P = -\Omega$. We take care of the vacuum pressure, taken numerically at $T=0.001$ and $\mu = 0$, by subtracting its contribution from the total pressure given as~\cite{Fuseau:2019zld},
\begin{equation}
    P = - \Big{(}\Omega(T, \mu) - \Omega(0.001,0) \Big{)}~.
\end{equation}
Now, the entropy density ($s$) and the energy density ($\epsilon$) are estimated as,
\begin{equation*}
\begin{split}
    s = - \frac{\partial \Omega}{\partial T} = \frac{\partial P}{\partial T}~,
\end{split}
\end{equation*}
\begin{equation*}
    \epsilon = - P + Ts~.
\end{equation*}
Furthermore, the speed of sound, $c_s$ and the isothermal compressibility, $\kappa_{\rm {T}}$ can be obtain as,
\begin{equation}
    c_{s}^{2} = \frac{\partial P}{\partial \epsilon} = \frac{\frac{\partial P}{\partial T}}{\frac{\partial \epsilon}{\partial T}}~,
\end{equation}
\begin{equation}
    \kappa_{T} = - \frac{1}{V} \frac{\partial V}{\partial P}~,
\end{equation}
where volume, $V = N/n$ with the total number of partons $N$ and the number density $n$. Plugging this into the above equation, we get
\begin{equation}
    \kappa_{T} = \frac{1}{n} \frac{\partial n}{\partial P} = \frac{1}{n} \frac{\frac{\partial n}{\partial \mu}}{\frac{\partial P}{\partial \mu}}~.
    \label{kTeqn}
\end{equation}

\subsection{Quark matter in finite magnetic field}
\label{formulation Bneq0}
Under the effect of an external magnetic field, the PNJL lagrangian changes as~\cite{Avancini:2011zz, Ferreira:2014kpa},
\begin{widetext}
\begin{equation}
\begin{split}
    {\mathcal L_{PNJL}^{B}} = \bar q(i \gamma^\mu D_\mu-\hat m)q + G_{B}(eB)\sum_{a=0}^{8} \left[ (\bar q\lambda^a q)^{2} + (\bar q i\gamma_5\lambda^a q)^{2} \right] - K\{{\rm det}[\bar q(1+\gamma_5)q] +\mbox{det}[\bar q(1-\gamma_5)q]\} \\ - \mathcal{U}( \Phi, \overline{\Phi}, T) - \frac{1}{4}F_{\mu\nu}F^{\mu\nu}
\end{split}
\end{equation}
\end{widetext}
The coupling of the quarks and the temporal effective gluon field with the external magnetic field is incorporated by adding the $\frac{1}{4}F_{\mu\nu}F^{\mu\nu}$ term, where $F_{\mu\nu} = \partial_{\mu}A_{\nu}^{EM} - \partial_{\nu}A_{mu}^{EM}$, and altering the covariant derivative as $D^{\mu} = \partial^{\mu}-i q_{f}A_{EM}^{\mu} - i A^{\mu}$, where $e_{f}$ is the electric charge of the quark of flavor $f$. The coupling constant $G$ is taken as, $G_{B}(eB)$, a function of $eB$ to incorporate the effect of an external magnetic field. Various forms of $G_{B}(eB)$ exist in literature~\cite{Farias:2014eca,  Farias:2016gmy, Yang:2017xdd, Li:2016dta, Ferreira:2013tba, Moreira:2020wau, Ferreira:2014kpa}. For our study, we use the form given in~\cite{Ferreira:2014kpa},
\begin{equation}
\label{GB}
    G_{B}(eB) = G\left( \frac{1 + a\zeta^{2} + b\zeta^{3}}{1 + c\zeta^{2} + d\zeta^{4}} \right)
\end{equation}
This form is obtained by reproducing the chiral transition temperature, $T_{C}^{\chi}(eB)$, obtained in lQCD~\cite{Bali:2011qj}. This form of coupling constant is able to yield qualitatively precise results for quark condensate values at a finite magnetic field.
where $a=0.0108805$, $b=-1.0133\times10^{-4}$, $c=0.02228$, $d=1.84558\times10^{-4}$, and $\zeta = eB/\Lambda^{2}_{QCD}$ with $\Lambda_{QCD} = 300$ MeV. Finally, the thermodynamic potential due to an external magnetic field changes as~\cite{Avancini:2011zz, Ferreira:2014kpa},
\begin{widetext}
\begin{equation}
\begin{split}
    \Omega (T, \mu, eB) = \mathcal{U}( \Phi, \overline{\Phi}, T)  + 2G_{B}(eB)\sum_{{f = u,d,s}} \sigma_{f}^2  -  4K\sigma_u \sigma_d \sigma_s +  \left(\sum_{{f = u,d,s}} \Omega_{f}^{vac} + \Omega_{f}^{med} + \Omega_{f}^{mag} \right)
\end{split}
\end{equation} 
\end{widetext}
where contributions from the vacuum, $\Omega^{vac}_{f}$, medium, $\Omega^{med}_{f}$, and magnetic field, $\Omega^{mag}_{f}$ which are given as~\cite{Menezes:2008qt, Menezes:2009uc, Ferreira:2014kpa},
\begin{align}
\label{eq15}
    \Omega_{f}^{vac} = -6\int_{\Lambda} \frac{d^{3}p}{(2\pi)^{3}}\sqrt{p^{2} + M_{f}^{2}},
\nn\\
\Omega_{f}^{med} = -T \frac{|e_{f}B|}{4\pi^{2}}\sum_{k}a(k)\int_{-\infty}^{\infty} dp_{z} (\mathcal{Z}_{\Phi}^{+} + \mathcal{Z}_{\Phi}^{-}),
\nn\\
 \Omega_{f}^{mag} = -\frac{3|e_{f}B|^{2}}{2\pi^2} \left[ \zeta'(-1, x_{f}) - \frac{1}{2}(x_{f}^{2} - x_{f}){\rm ln}(x_{f}) + \frac{x_{f}^{2}}{4} \right].
\end{align}
Here, the contribution from the vacuum ($\Omega_{f}^{vac}$) is regularized with a three-momentum cutoff scheme. In this scheme, the particles above a sharp momentum cutoff ($\Lambda = 602.3~\rm{MeV}$) are excluded. Thus, allowing the vacuum term to restrict itself to a finite value.
For $\Omega_{f}^{mag}$, the authors in Ref.~\cite{Menezes:2008qt} have taken care of the divergent part by using the standard dimensional regularization formula. Moreover, the contribution coming from medium ($\Omega_{f}^{med}$) is a converging function, and thus no regularization scheme has been applied for this part. However, in literature, there exist a few studies where the authors have considered the $\Lambda$ cutoff for $\Omega_{f}^{med}$ part as well~\cite{Costa:2009ae, Xue:2021ldz}. It has been observed in these studies that considering the medium contributions without any three-momentum cutoff yields better results for quantities such as pressure, energy density, specific heat, and speed of sound as well as deconfinement temperature.
The details of the three-momentum cutoff and other possible regularization schemes have been explained in Ref.~\cite{Klevansky:1992qe, Menezes:2008qt, Kohyama:2015hix}  

In Eq.~\ref{eq15}, ${Z}_{\Phi}^{+}$ and ${Z}_{\Phi}^{-}$ are the partition function density given as\cite{Hansen:2006ee, Torres-Rincon:2017zbr},
\begin{equation*}
\begin{split}
    {Z}_{\Phi}^{+} = {\rm ln}\Big{[} 1 + 3 \left(\Phi + \overline{\Phi} e^{-\beta(E_{f}-\mu_{f})}  \right) e^{-\beta(E_{f}-\mu_{f})} \\ +  e^{-3\beta(E_{f}-\mu_{f})} \Big{]} \\
    {Z}_{\Phi}^{-} = {\rm ln}\Big{[} 1 + 3 \left( \overline{\Phi} + \Phi e^{-\beta(E_{f}+\mu_{f})} \right) e^{-\beta(E_{f}+\mu_{f})}  \\ +  e^{-3\beta(E_{f}+\mu_{f})} \Big{]}~.
    \end{split}
\end{equation*}
$x_{f} = M_{f}^{2}/(2|e_{f}| B)$, $\zeta (z,x) $ is the Riemann-Hurwitz zeta function, and $\zeta'(-1, x_{f}) = d\zeta(z,x_{f})/dz\mid_{z=-1}$ can be expressed as \cite{ Pradhan:2021vtp, webiste:2023aa}, 
\begin{equation*}
\begin{split}
    \zeta'(-1, x_{f}) = \frac{1}{12} - \frac{x^{2}_{f}}{4} + (\frac{1}{12} - \frac{x_{f}}{2} + \frac{x^{2}_{f}}{2}){\rm log}(x_{f}) + \mathcal{O}(x^{-2}_{f})
\end{split}
\end{equation*}
In the presence of a magnetic field, the dispersion relation of quarks will be modified due to the Landau quantization as,
\begin{equation}
    E_{f} = \sqrt{p^{2} + M_{f}^{2} + 2k|e_{f}|B}~.
\end{equation}
Minimizing the thermodynamic potential with respect to $\Phi$, $\bar{\Phi}$ and $\sigma_{f}$, we get a set of coupled equations to be solved simultaneously~\cite{Fu:2007xc, Costa:2008dp},

\begin{equation*}
    \frac{\partial \Omega}{\partial \sigma_f} = 0,   ~\frac{\partial \Omega}{\partial \Phi} = 0,   ~\frac{\partial \Omega}{\partial \bar{\Phi}} = 0
\end{equation*}
Further, the condensates split under the effect of an external magnetic field as,
\begin{equation*}
    \sigma_{f} = \sigma_{f}^{vac} + \sigma_{f}^{med} + \sigma_{f}^{mag}
\end{equation*}
where $\sigma_{f}^{vac}$, $\sigma_{f}^{med}$, and $\sigma_{f}^{mag}$ are the contributions from vacuum, medium and the magnetic field respectively, which are given as~\cite{Ferreira:2013oda, Wang:2020mww, Menezes:2008qt, Farias:2016gmy, Torres-Rincon:2017zbr},
\begin{align}
    \sigma_{f}^{vac} = -6 \int_{\Lambda}\frac{d^{3}p}{(2\pi)^{3}}\frac{M_{f}}{E_{f}}~,
\nn\\
    \sigma_{f}^{med} = \frac{3(|e_{f}B|)^{2}}{4\pi^{2}}\sum_{k}a(k)\int_{-\infty}^{\infty} \frac{dp_{z}}{E_{f}} (f_{\Phi}^{+} + f_{\Phi}^{-})~,
\nn\\
    \sigma_{f}^{mag} = -\frac{M|e_{f}|BN_{c}}{2\pi^{2}}\Big{[} {\rm ln \Gamma}(x_{f}) - \frac{1}{2}{\rm ln}(2\pi) \nn\\
    + x_{f} -\frac{1}{2}(2x_{f} - 1){\rm ln}(x_{f}) \Big{]}~,
\end{align}

Any thermodynamic quantity can be derived from the thermodynamic potential $\Omega$. In the presence of a magnetic field, pressure becomes anisotropic due to the Landau quantization. Pressure along the direction of the magnetic field is longitudinal pressure ($P^\parallel$) and in the direction perpendicular to the magnetic field it is transverse pressure ($P^\perp$), and can be expressed as\cite{Chu:2016rpw, Yang:2021rdo},
\begin{align}
    P^{\parallel} = -\Omega - \frac{(eB)^{2}}{2}~,
\nn\\
    P^{\perp} = -\Omega - eB \mathcal{M} + \frac{(eB)^{2}}{2}~,
\end{align}
where $\mathcal{M}$ is the magnetization,
\begin{equation}
    \mathcal{M} = - \Bigg{(} \frac{\partial \Omega}{\partial eB} \Bigg{)}_{\mu}~.
\end{equation}
The normalized pressure is taken in a way that it vanishes at vacuum~\cite{Menezes:2008qt, Ferreira:2013oda, Li:2016dta, Farias:2016gmy},
\begin{equation*}
    P^{\parallel}_{N} = P^{\parallel}(T, \mu, eB) - P^{\parallel}(0.001, 0, eB)
\end{equation*}
\begin{equation*}
    P^{\perp}_{N} = P^{\perp}(T, \mu, eB) - P^{\perp}(0.001, 0, eB)~.
\end{equation*}
In the weak magnetic field limit, for simplicity, we can take $P^{\perp}_{N} \simeq P^{\parallel}_{N} = P_{N}$. Using the Euler thermodynamic relation, energy density can be written as,
\begin{equation}
    \epsilon = - P_{N} + Ts + \mathcal{M}eB
\end{equation}
The speed of sound also becomes anisotropic in the presence of a magnetic field due to changes in pressure in the longitudinal and transverse directions. The anisotropic squared speed of sound reads as~\cite{Borsanyi:2012cr, Piattella:2013wpa},
\begin{align}
    c_{s}^{2 \parallel(\perp)} = \Bigg{(} \frac{d P^{\parallel(\perp)}_{N}}{d \epsilon} \Bigg{)}_{s/n}~, \\ 
    c_{s}^{2 \parallel(\perp)} =  \frac{\frac{\partial P^{\parallel (\perp)}_{N}}{\partial T} + \frac{\partial P^{\parallel (\perp)}_{N}}{\partial eB}\frac{deB}{dT}}{\frac{\partial \epsilon}{\partial T} + \frac{\partial \epsilon}{\partial eB}\frac{deB}{dT}}~,
\end{align}
where 
\begin{equation*}
\frac{deB}{dT} = \frac{s\frac{\partial n}{\partial T} - n\frac{\partial s}{\partial T}}{n\frac{\partial s}{\partial eB} - s\frac{\partial n}{\partial eB}}.
\end{equation*}
Isothermal compressibility can be expressed as in Eq.~(\ref{kTeqn}) for the vanishing magnetic field. For a system with a finite magnetic field, the pressure term in the denominator splits isothermal compressibility into  longitudinal and transverse components, which can be expressed as \cite{Yang:2021rdo},
\begin{align}
    \kappa_{T}^{\parallel} = \frac{1}{n} \frac{\frac{\partial n}{\partial \mu}}{\frac{\partial P^{\parallel}}{\partial \mu}} = \frac{1}{n^{2}}{\frac{\partial n}{\partial \mu}}~.
\nn\\
    \kappa_{T}^{\perp} = \frac{1}{n} \frac{\frac{\partial n}{\partial \mu}}{\frac{\partial P^{\perp}}{\partial \mu}} = \frac{1}{n}{\frac{\partial n}{\partial \mu}}\Bigg{(}\frac{1}{n - \frac{\partial \mathcal{M}}{\partial \mu}B} \Bigg{)}~.
\end{align}

\section{Results and Discussion}
\label{sec result}
\begin{figure*}
    \centering
    \includegraphics[width=0.45\linewidth]{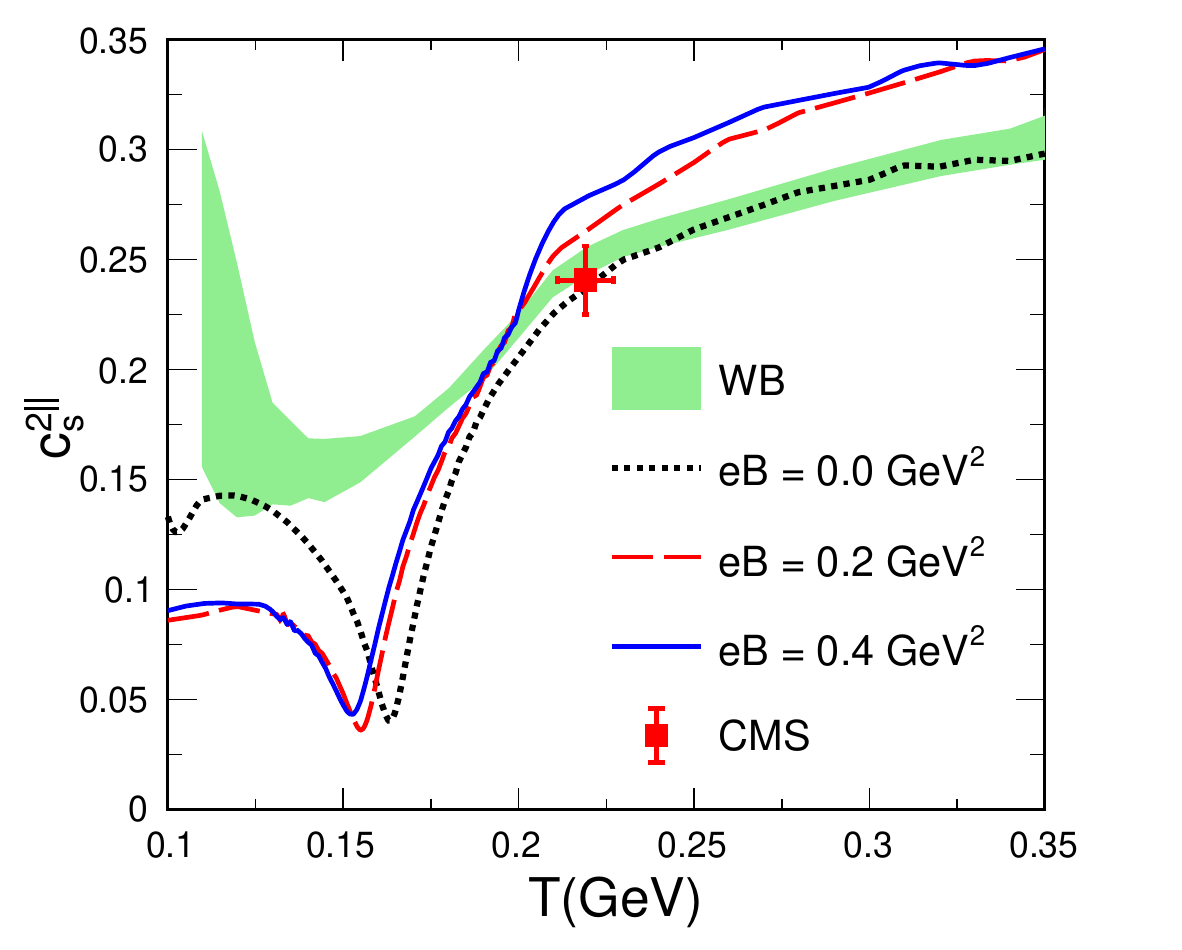}
    \includegraphics[width=0.45\linewidth]{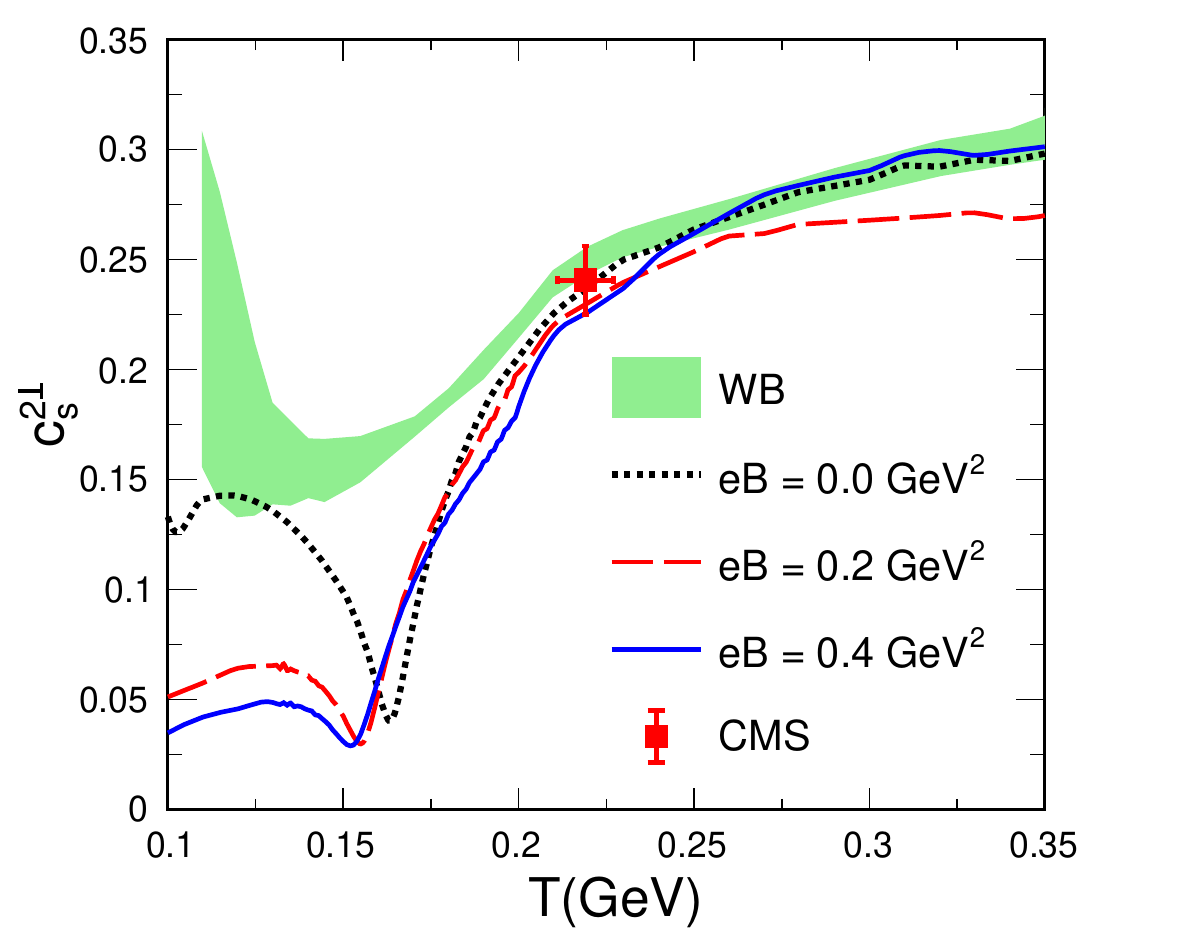}
        \caption{Squared speed of sound's longitudinal ($c_s^{2 \parallel}$) (left) and transverse ($c_s^{2 \perp}$) (right) components against temperature ($T$)}
    \label{figCs2}
\end{figure*}
\begin{figure*}
    \centering
    \includegraphics[width=0.3\linewidth]{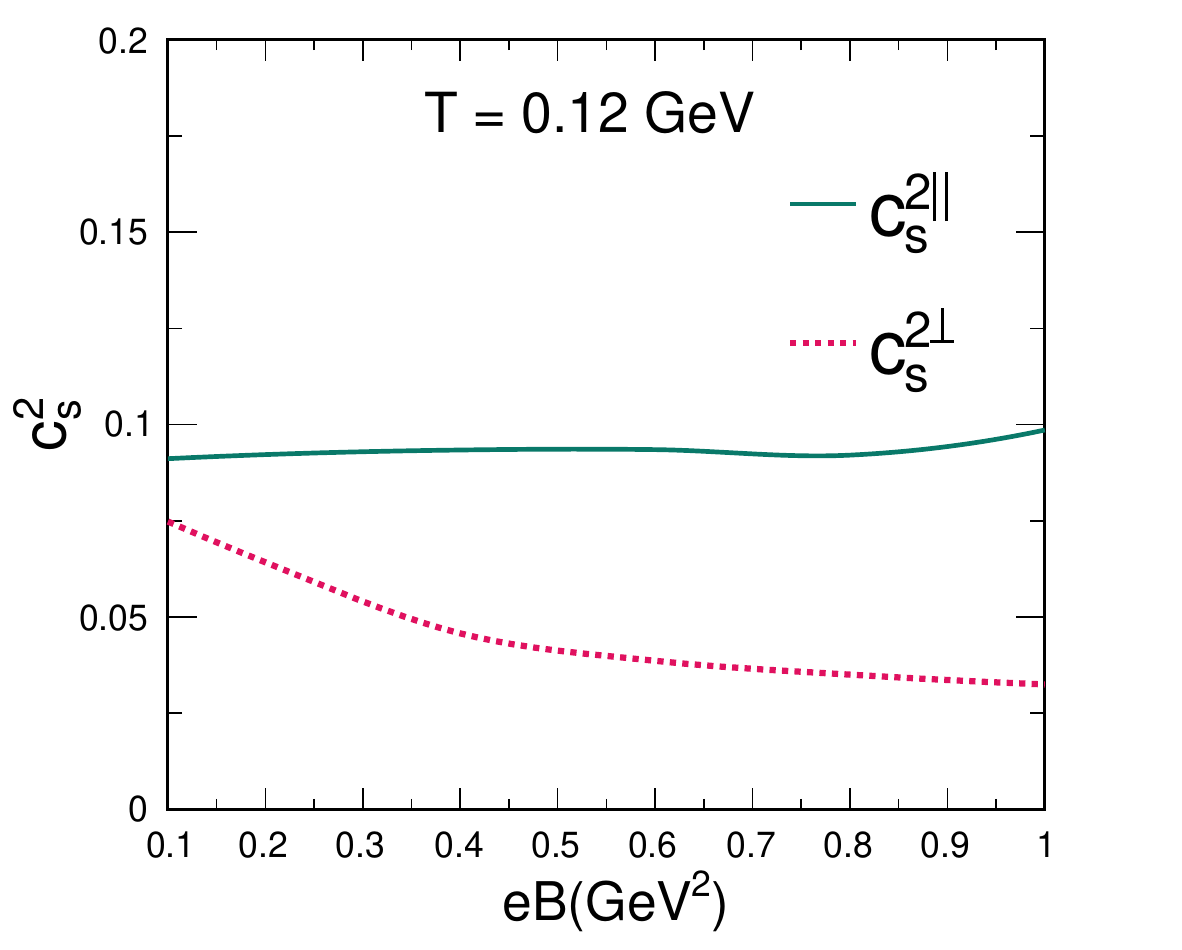}
    \includegraphics[width=0.3\linewidth]{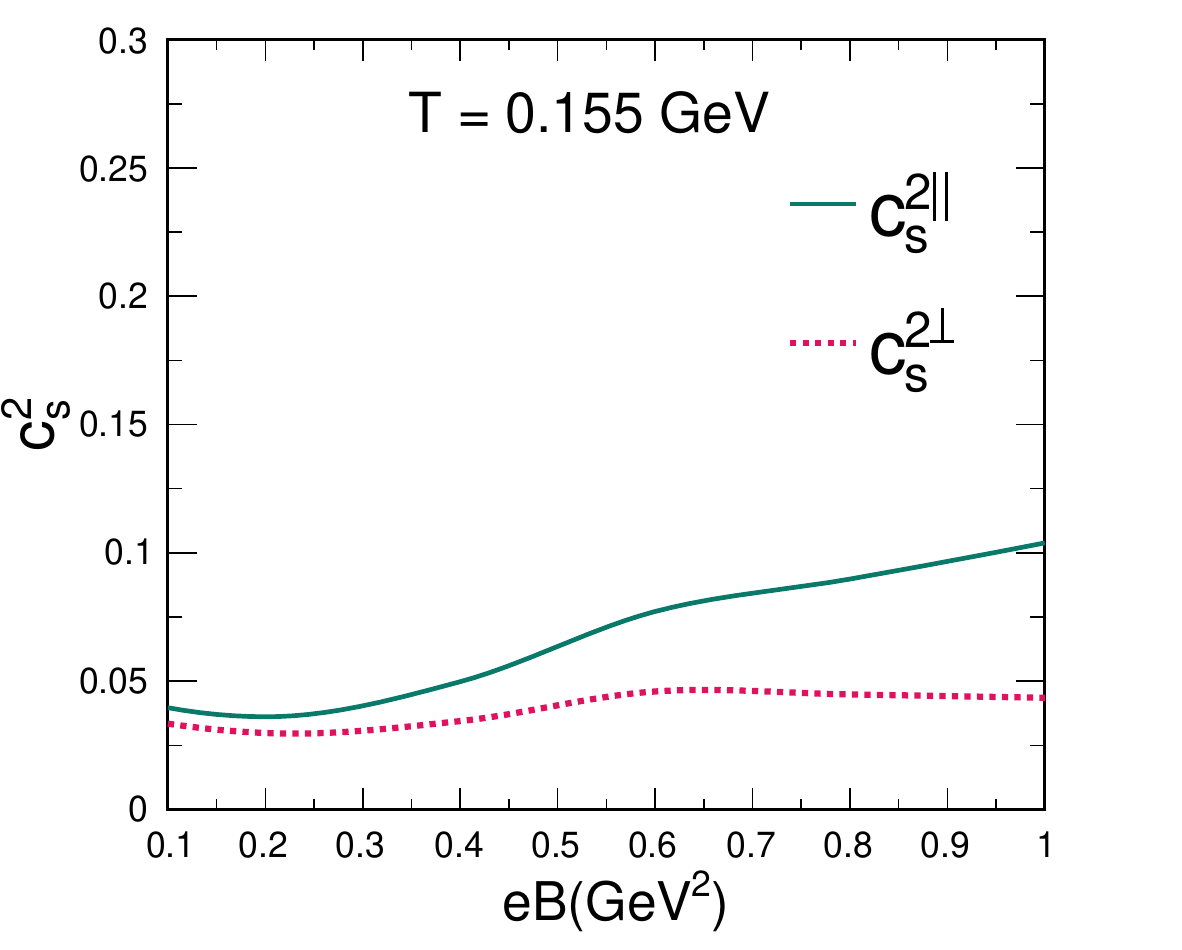}
    \includegraphics[width=0.3\linewidth]{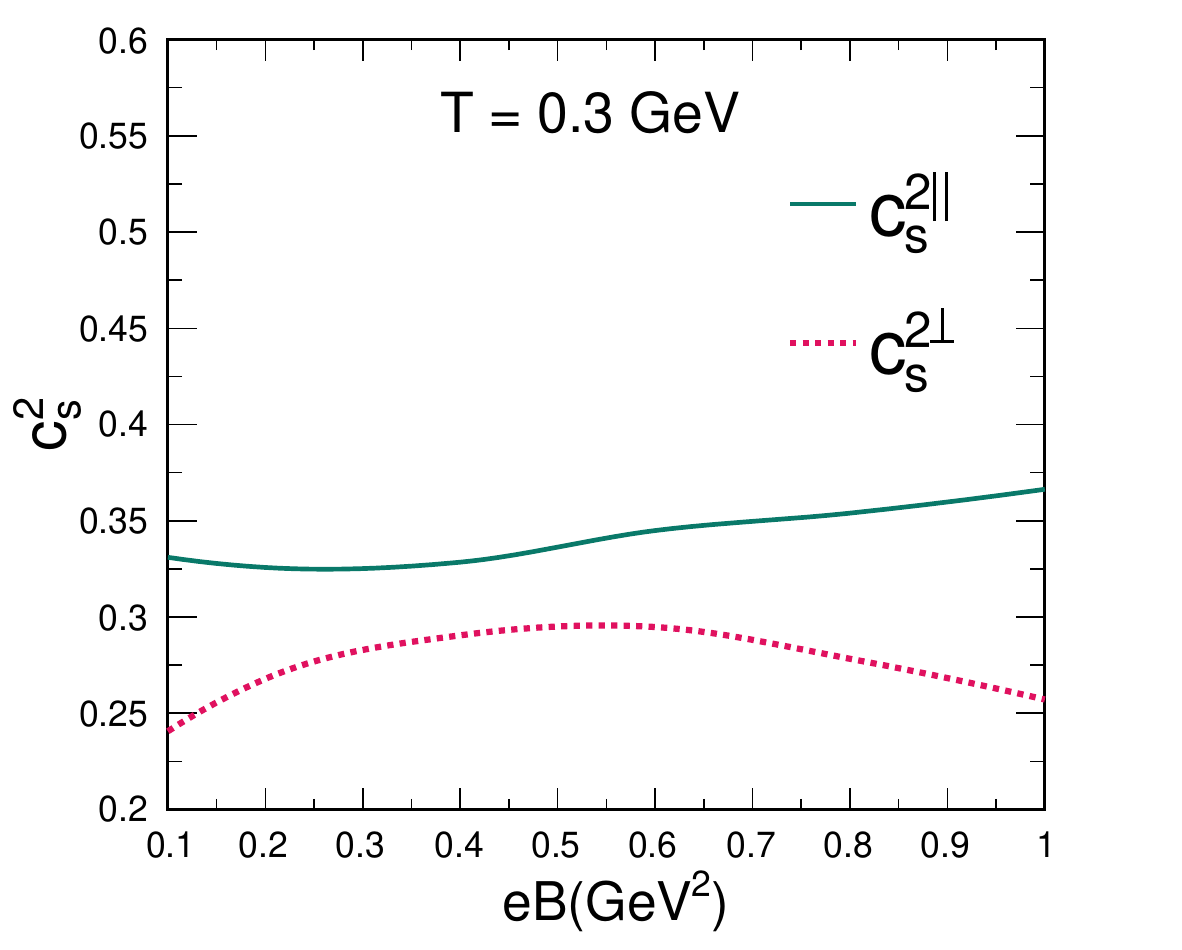}
        \caption{Squared speed of sound ($c_s^{2 \parallel}$ and $c_s^{2 \perp}$) against magnetic field ($eB$) for different temperature; $T = 0.12$~GeV (left), $T = 0.155$~GeV (middle), and $T = 0.3$~GeV (right)}.
    \label{figCs2B}
\end{figure*}
\begin{figure*}
    \centering
    \includegraphics[width=0.45\linewidth]{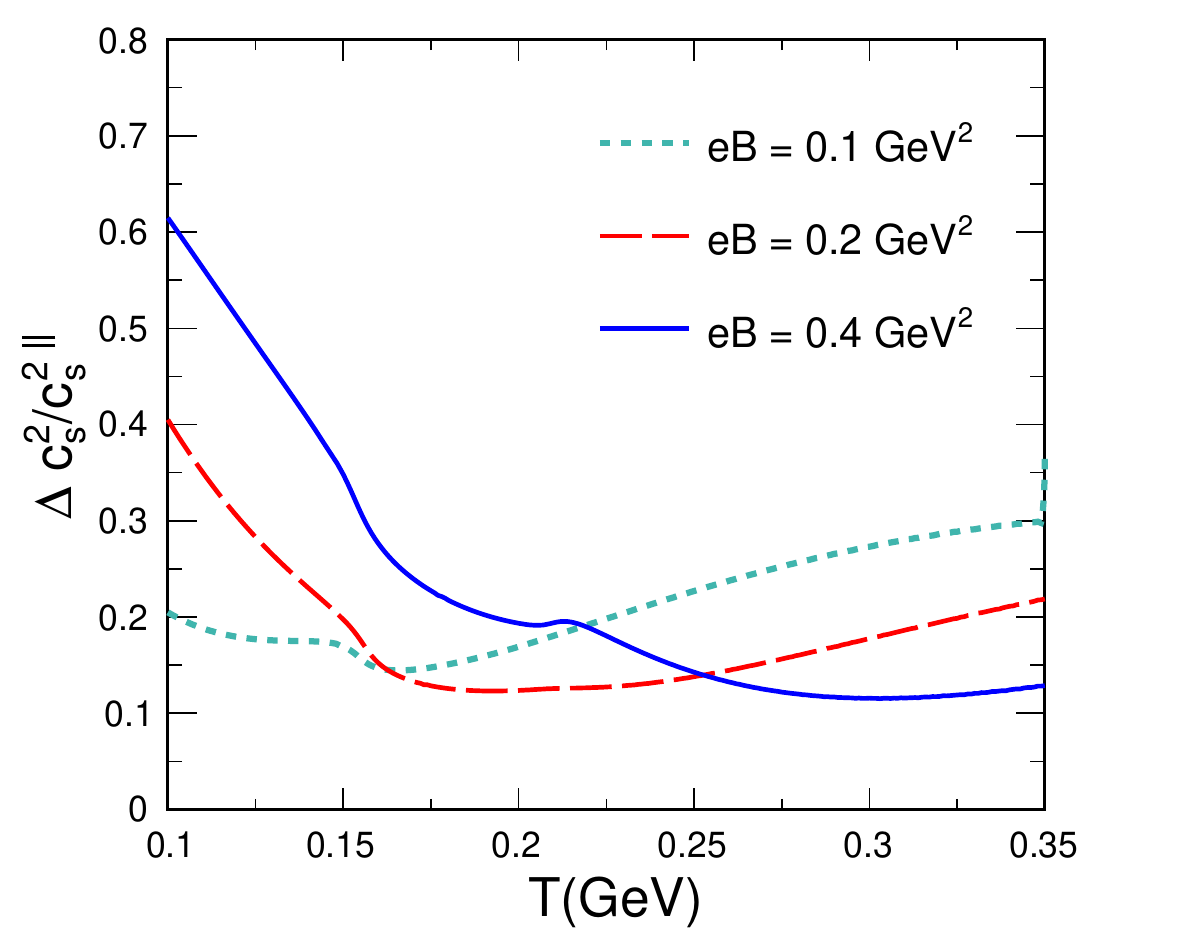}
    \includegraphics[width=0.45\linewidth]{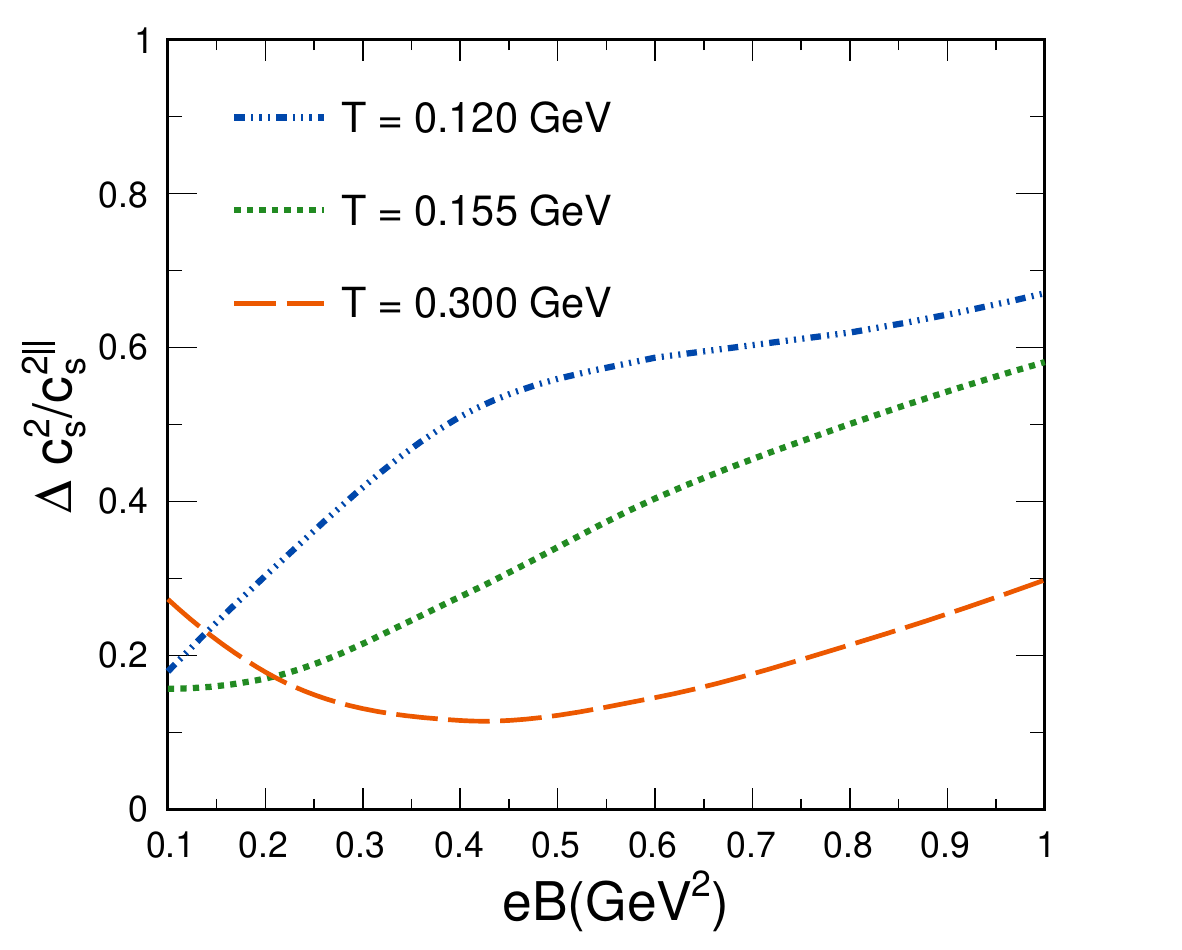}
        \caption{Anisotropy in squared speed of sound ($\Delta c_s^2/c_s^{2\parallel} $) against $T$ (left) and $eB$ (right). Here, $\Delta c_s^2 = c_s^{2\parallel} - c_s^{2\perp}$.}
    \label{fig speed aniso}
\end{figure*}

In the presence of a magnetic field, $c_s^2$ becomes anisotropic. It signifies that any disturbance in the medium will not travel at the same speed in all directions in the presence of an external magnetic field. In Fig.~\ref{figCs2}, longitudinal ($\parallel$) (left) and transverse ($\perp$) (right) components of $c_s^2$ are plotted against temperature for different values of magnetic fields; $eB = 0.0~\rm GeV^{2}$ (black dotted line), $eB = 0.2~\rm GeV^{2}$ (red dashed line), and $eB = 0.4~\rm GeV^{2}$ (solid blue line). Lattice estimation at the zero magnetic fields is also presented here from WB group~\cite{BORSANYI201499}. We also compare with the latest CMS estimation of speed of sound in a central Pb-Pb collision at $\sqrt{s_{NN}} = 5.02~\rm{TeV}$~\cite{Bernardes:2023nnf}, where a relation between $\partial{P}/\partial{\epsilon}$ and experimental observables such as $N_{ch}$ and $\langle p_{T} \rangle$ has been established to obtain a $c_{s}^{2}$ estimation from the CMS data. The study is done for head-on collision where no magnetic field is expected and as a result,  our estimations, for $eB = 0.0~\rm GeV^{2}$ case, matches with the results taken from the CMS data. Moreover, it will be interesting to perform similar studies for peripheral heavy-ion collisions where strong transient magnetic field is expected. Estimating speed of sound for such collisions will shed light on the effect of magnetic field in speed of sound. At zero magnetic field, the PNJL result agrees with lattice estimation at the high temperature. The dip around transition temperature ($T_c$) in the PNJL result is the consequence of deconfinement phase transition which is absent in NJL estimations~\cite{Deb:2016myz}. With the increasing magnetic field, the dip shifts towards lower temperatures. This signifies that with the increasing magnetic field, the deconfinement temperature decreases, which agrees with the lattice results obtained in Ref.~\cite{Braguta:2019yc}.

In a massless non-interacting (ideal) gas, the speed of sound is estimated to be $1/3$~\cite{PARDY201370}. In Fig.~\ref{figCs2}, we observe that as the temperature increases, chiral symmetry is restored, and $c_s^2$ approaches the ideal gas limit. We observe an interesting trend of $c^{2}_{s}$ near the transition temperature.
For $T<T_c$ (or below the dip in the speed of sound), $c_s^2$ decreases with increasing magnetic field, i.e., the speed of sound shows an order as $(c_{s}^{2})^{B=0} > (c_{s}^{2})^{B=0.2} > (c_{s}^{2})^{B=0.4}$. However, slightly above $T_c$, the speed of sound is higher at a larger magnetic field, i.e., the speed of sound shows an order as $(c_{s}^{2})^{B=0.4} > (c_{s}^{2})^{B=0.2} > (c_{s}^{2})^{B=0}$. This behavior is more prominent in the parallel component of $c_s^2$. This behavior is a consequence of magnetic catalysis (MC) and inverse magnetic catalysis (IMC) phenomena. Due to magnetic catalysis, at low temperatures, the quark condensate is enhanced in the presence of a magnetic field, and as a result, the dynamical mass of the quarks is directly proportional to the magnetic field strength. However, due to inverse magnetic catalysis, beyond $T_{c}$, the quark condensate is reduced with increasing magnetic field and the mass is inversely proportional to the external magnetic field. The same behaviour can be seen for the speed of sound near $T_{c}$, thus hinting at the fact that speed of sound is sensitive to mass ordering. These phenomena have been widely explored in literature~\cite{Bali:2012zg, Ferreira:2014kpa, Farias:2016gmy}.

In Fig.~\ref{figCs2B}, $c_s^2$ is plotted against $eB$ for three different values of temperatures, such that $T = 0.12$~GeV (left) represents the confined or hadronic medium, $T = 0.155$~GeV (middle) is around $T_c$, and $T = 0.3$~GeV (right) represents the deconfined or partonic medium.
$\parallel$ and $\perp$ components are represented by a solid green and red dotted line, respectively. $c_s^2$ shows a non-monotonic behavior with respect to $eB$. $\parallel$ component is almost independent of the magnetic field because, along the magnetic field, direct contribution from Lorentz force vanishes. In the confinement zone, the effect of the magnetic field is large in the $\perp$ component. We observe a decrease in the transverse component of the speed of sound with an increase in magnetic field strength. Around the transition temperature, the effect of the magnetic field on both of the components is not significant. However, there is a steady rise for the longitudinal component at a strong magnetic field.  On the right, for the partonic phase, we have both weak and strong field regions. In the weak field region, we observe an increasing and then decreasing behavior that could be attributed to the interplay of decreasing coupling constant with an increase in $eB$ and dependence of dynamical mass on the magnetic field strength.

To better understand the anisotropy in the components of $c_s^2$ due to magnetic field, we have plotted $\Delta c_s^2/c_s^{2\parallel}$ with $T$ for three different values of magnetic field ($eB = 0.1,~0.2,~0.4~\rm{GeV}^{2}$) in Fig.~(\ref{fig speed aniso}) (left). On the right panel, we have plotted the same against $eB$ for different temperatures ($T = 0.12,~0.155,~0.3$~GeV). Here, $\Delta c_s^2 = c_s^{2\parallel} - c_s^{2\perp}$ represents normalized anisotropy in the components of $c_s^2$. In other words, Fig.~(\ref{fig speed aniso}) represents the deviation of $\perp$ component from $\parallel$ component in the presence of magnetic fields. In the left panel, anisotropy is larger at a low temperature and decreases as temperature increases. For $eB=0.2,~0.4~\rm{GeV}^{2}$, anisotropy is around $20\%$ to $60\%$, respectively, at $T = 0.1$~GeV. The degree of anisotropy decreases with temperature and obtains minima near the transition temperature. The position of the minima changes with magnetic field strength. With an increase in temperature, the deviation increases as well. Moreover, we observe two kink-like structures, one near $150-160~\rm{MeV}$ and a second one near $200-220~\rm{MeV}$. The first one is due to the effect of deconfinement temperature, and the second one emerges due to the change in the system near the chiral transition temperature. In Ref.~\cite{Ferreira:2014exa}, the authors have explicitly shown the effect of deconfinement transition temperature (maxima of the Polyakov loop susceptibility) and chiral transition temperature (maxima of the quark condensate susceptibility) in the speed of sound and specific heat.

The figure on the right shows that anisotropy increases as the magnetic field increases. Anisotropy is smaller at low $eB$ for all three temperatures and increases gradually with increasing $eB$. Therefore, from both figures, we can conclude that anisotropy is higher in the strong field zone, i.e., low $T$ and high $eB$. With decreasing magnetic field, anisotropy diminishes, and $\parallel$ and $\perp$ components merge to the isotropic component of $c_s^2$ as soon as the magnetic field vanishes.

\begin{figure}
    \centering
\includegraphics[scale = 0.45]{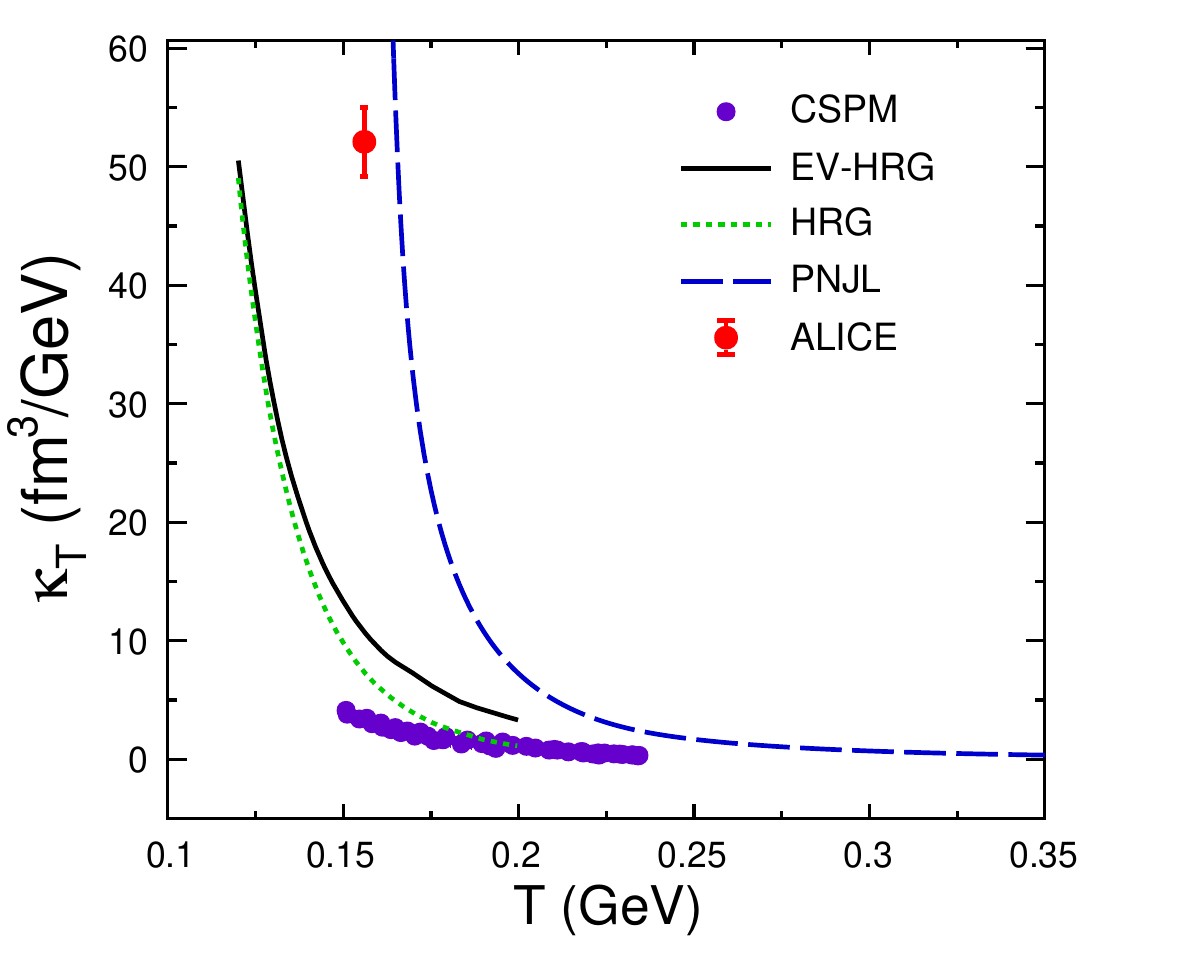}
    \caption{Isothermal compressibility ($\kappa_{T}$) against temperature ($T$) at zero magnetic field. A comparison of this work (PNJL) with different models (HRG, EVHRG~\cite{Khuntia:2018non}, and CSPM~\cite{Sahu:2020nbu}), and experimental results from ALICE~\cite{ALICE:2021hkc}.}
    \label{Comparison}
    \label{kTcomp}
\end{figure}

\begin{figure*}
    \centering
    \includegraphics[width=0.45\linewidth]{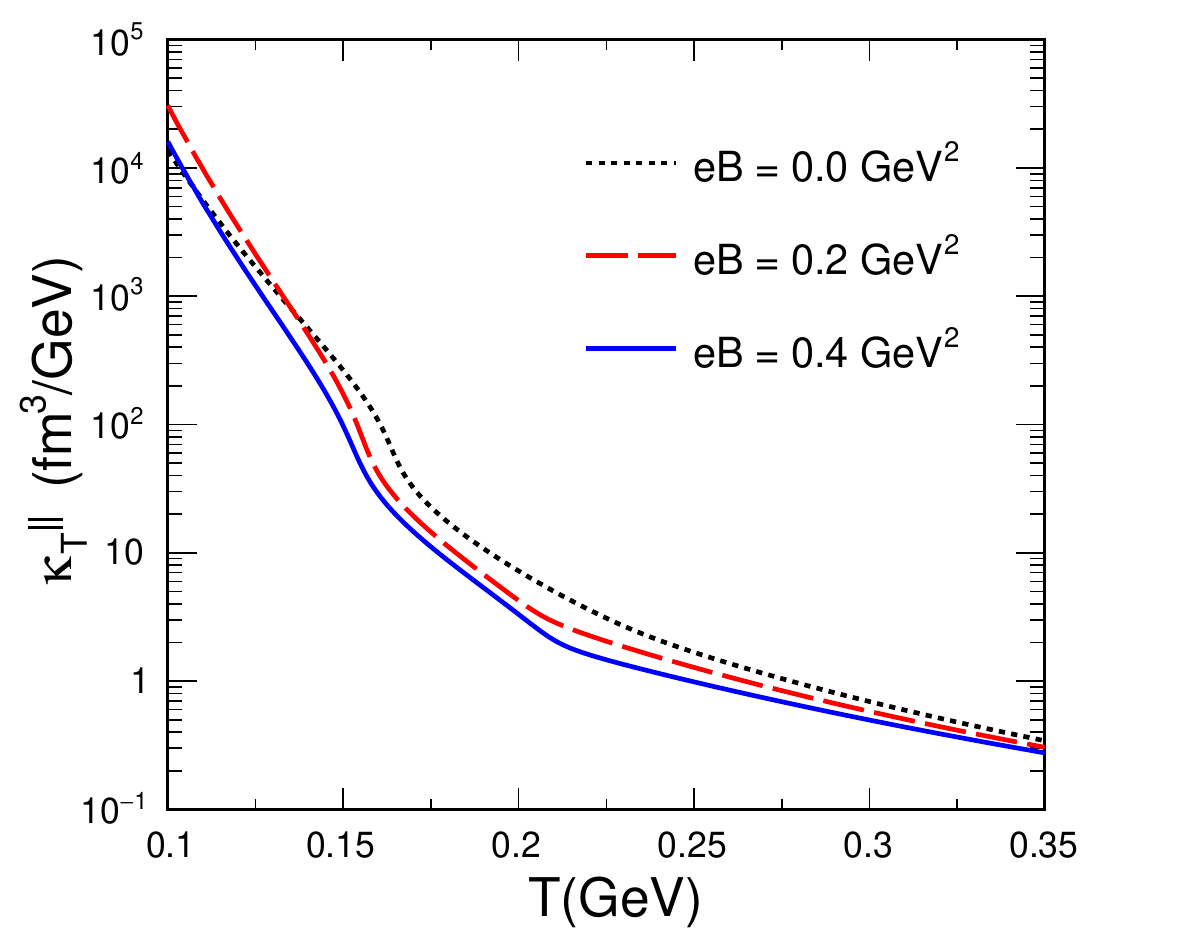}
    \includegraphics[width=0.45\linewidth]{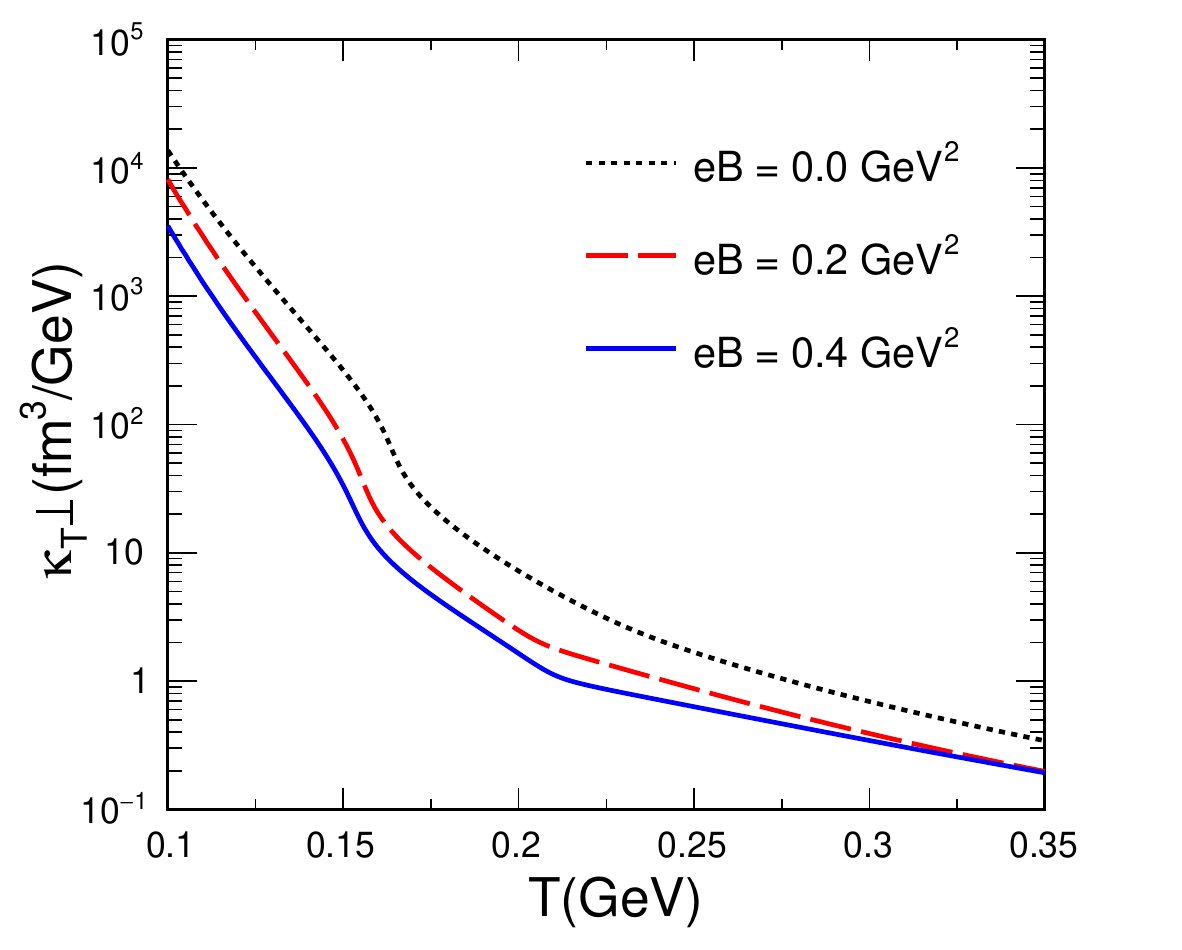}
    \caption{Isothermal compressibility: longitudinal component ($\kappa_T^\parallel$) (left) and transverse component ($\kappa_T^\perp$) (right) against temperature ($T$) for different value of magnetic field ($eB$).}
        \label{ICT}
\end{figure*}

\begin{figure*}
    \centering
    \includegraphics[width=0.3\linewidth]{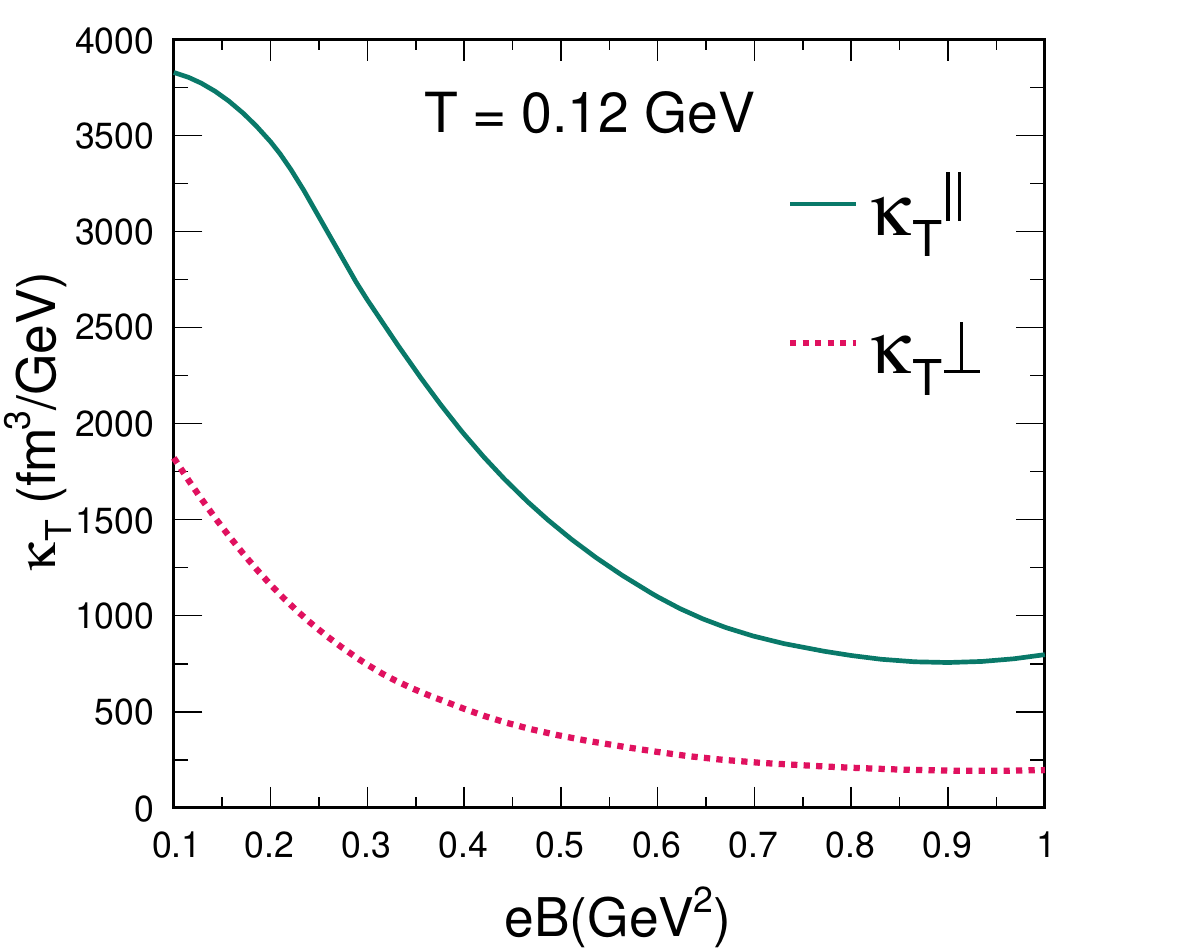}
    \includegraphics[width=0.3\linewidth]{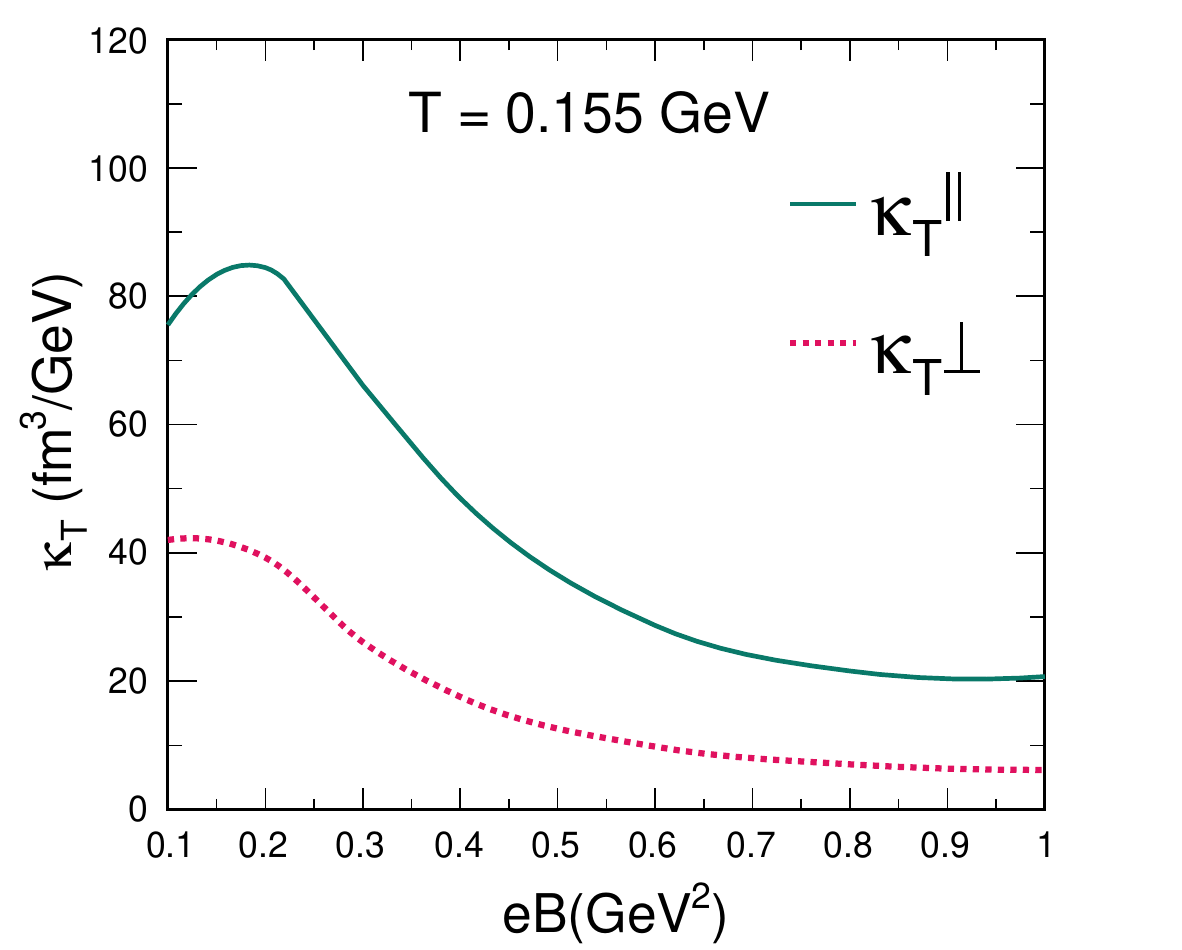}
    \includegraphics[width=0.3\linewidth]{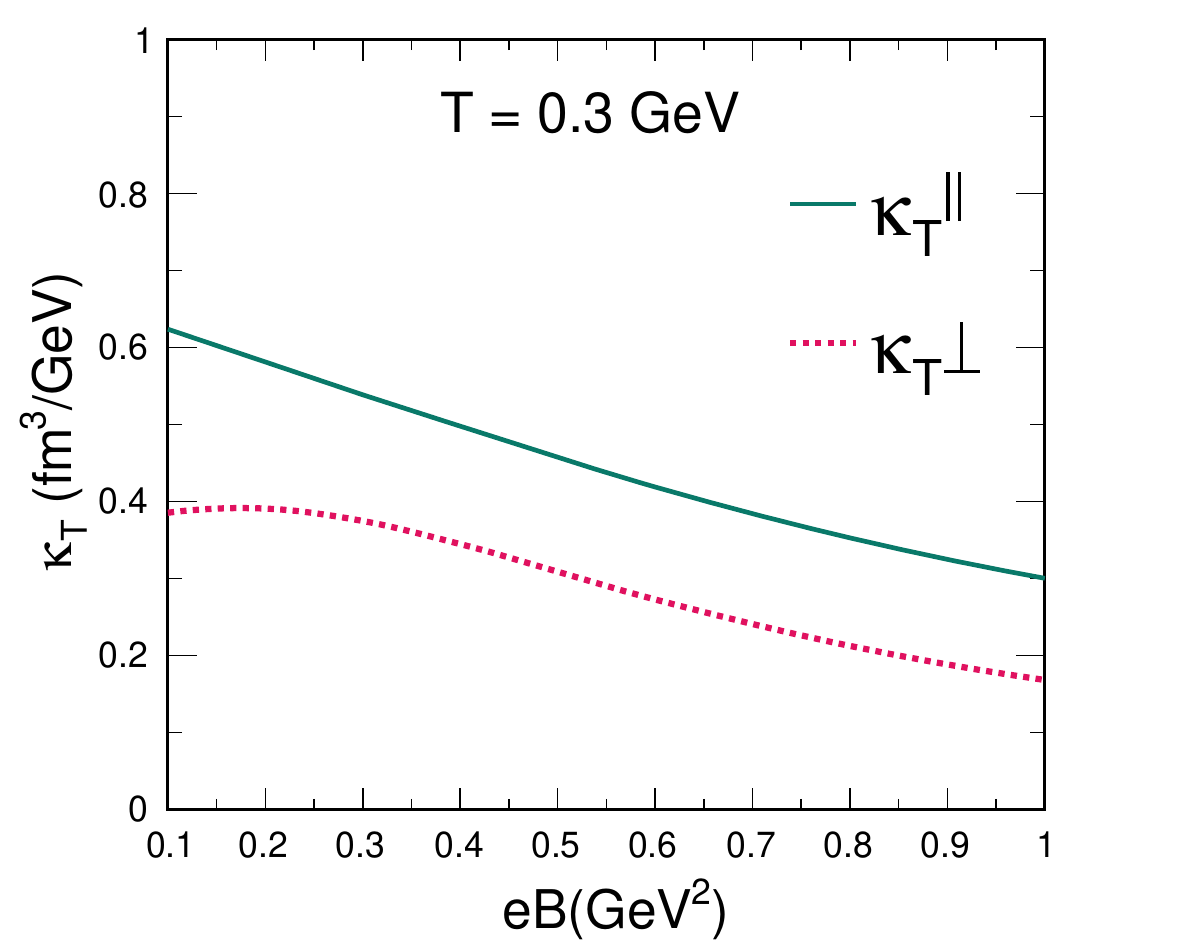}
    \caption{$\kappa_T^\parallel$ and $\kappa_T^\perp$ against $eB$ at $T = 0.12$~GeV (left), $T = 0.155$~GeV (middle), and $T = 0.3$~GeV (right).}
    \label{ICB}
\end{figure*}

In Fig.~\ref{Comparison}, $\kappa_T$ of QCD matter is plotted against temperature for QGP using the PNJL model. We compare our results with experimental results and different models. Experimentally, the isothermal compressibility is estimated using charged particle multiplicity fluctuations in Pb-Pb collisions at $\sqrt{s_{NN}}=2.76$ TeV~\cite{ALICE:2021hkc}. However, this estimation requires the information of temperature and volume of the system at chemical freeze-out which is taken from the statistical hadronization model. Results from hadronic models HRG (green dotted line), EV-HRG (solid black line)~\cite{Khuntia:2018non}, and partonic models CSPM (solid violet circle)~\cite{Sahu:2020nbu}, and the PNJL (blue dash line) in the absence of magnetic field are presented for comparison. $\kappa_T$ indicates stiffness in the equation of state (EoS). The smaller the value of $\kappa_T$, the stiffer the EoS is. Here, $\kappa_T$ decreases as $T$ increases, which means at very high $T$, the QCD matter is highly incompressible. This makes the QGP matter almost perfect fluid. In the absence of magnetic fields, $\kappa_T$ is isotropic. The EoS is the same in all directions. As we introduce the magnetic field, $\kappa_T$ becomes anisotropic, and we get two independent components. Along the magnetic field, EoS is defined by $\kappa_T^\parallel$ component, and in the plane perpendicular to the magnetic field, it is $\kappa_T^\perp$.

In Fig.~(\ref{ICT}), we have plotted $\kappa_T^\parallel$ (left) and $\kappa_T^\perp$ (right) as a function of $T$ for $eB = 0, ~0.2,~0.4~\rm{GeV}^2$. $\kappa_T$ is a proxy to the phase transition. For the baryon-free ($\mu_{\rm{B}} =0$) QCD matter, our results show a smooth transition from deconfinement to confinement temperature zone. Even in the presence of a magnetic field, both the components show smooth transition or crossover phase transition. A similar behavior as Fig.\ref{fig speed aniso} of kink-like structure can be seen here as well, where the kinks arise as a result of the effect of deconfinement temperature and chiral transition temperature in the medium.
However, one recent study~\cite{Yang:2021rdo} in NJL model discovered a first-order phase transition in $\kappa_T^\parallel$ component in the presence of a magnetic field at very low temperature and finite baryon chemical potential, which may not be realized in RHIC or LHC energies. Here, we see that the magnetic field reduces the compressibility, and the effect is nearly independent of $T$.

In Fig.~(\ref{ICB}), we have plotted $\kappa_T^\parallel$ (green solid line) and $\kappa_T^\perp$ (red dotted line) as a function of $eB$ for $T = 0.12$~GeV (left), $T = 0.155$~GeV (middle), and $T = 0.3$~GeV (right). At any fixed temperature, both the components decrease with increasing magnetic field. In the middle figure, where $T$ is around $T_c$, a peak-like structure is seen in $\kappa_T^\parallel$. The peak emerges as a result of deconfinement in the system. Similarly, a kink-like structure appears when it is plotted against temperature (Fig.~\ref{ICT}), and the kink moves towards a lower temperature for a higher magnetic field. Here, for $eB \simeq 0.2~\rm{GeV^{2}}$, the deconfinement temperature is around $T\simeq0.155~\rm{GeV}$, which gives rise to the observed peak. For a lower temperature, the peak will shift towards a higher $\rm{eB}$ value.
In the low-temperature regime, saturation in $\kappa_{T}$ can be observed for both parallel and perpendicular components. The value of the $\perp$ component for $\kappa_T$ is smaller compared to the $\parallel$ component. This signifies that along the magnetic field, EoS is stiffer. 
\begin{figure*}
    \centering
    \includegraphics[width=0.45\linewidth]{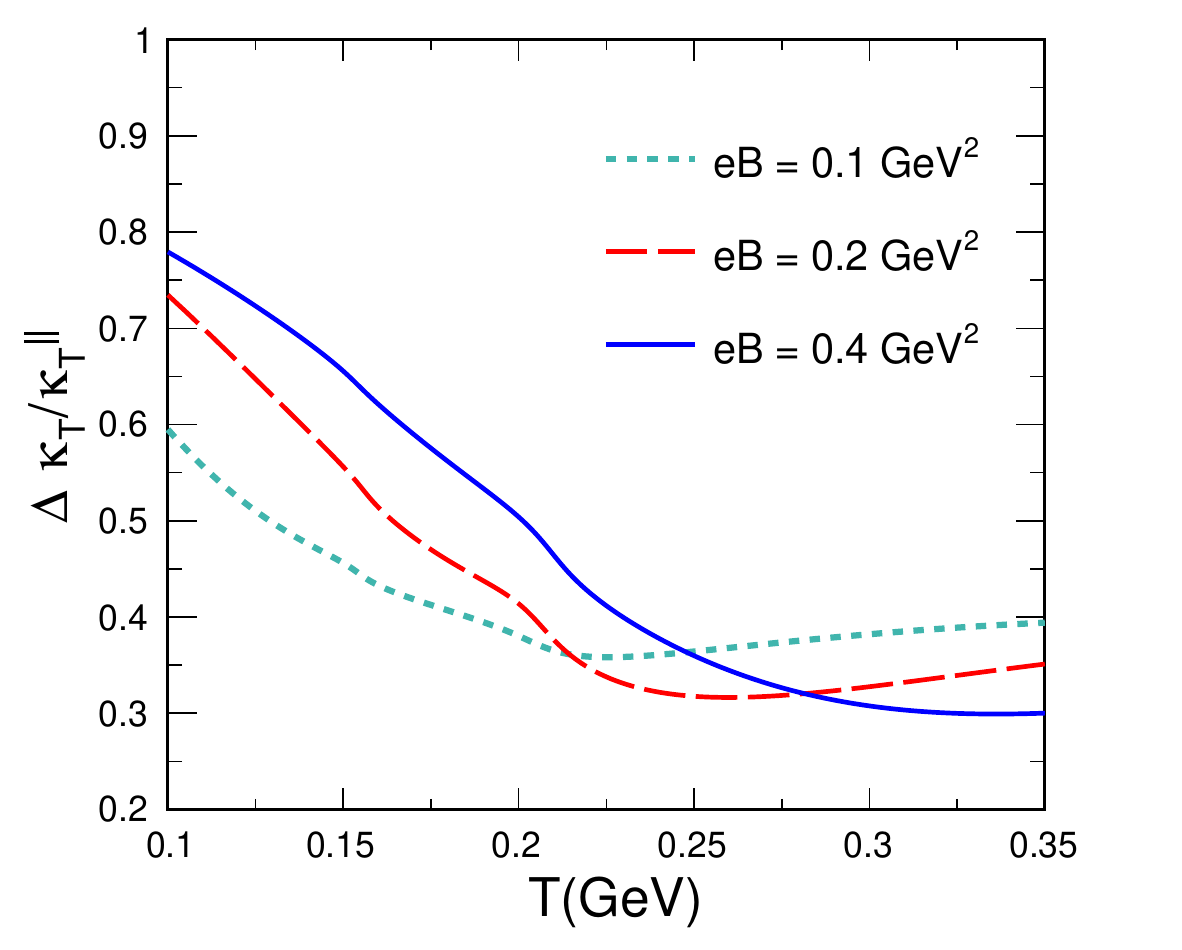}
    \includegraphics[width=0.45\linewidth]{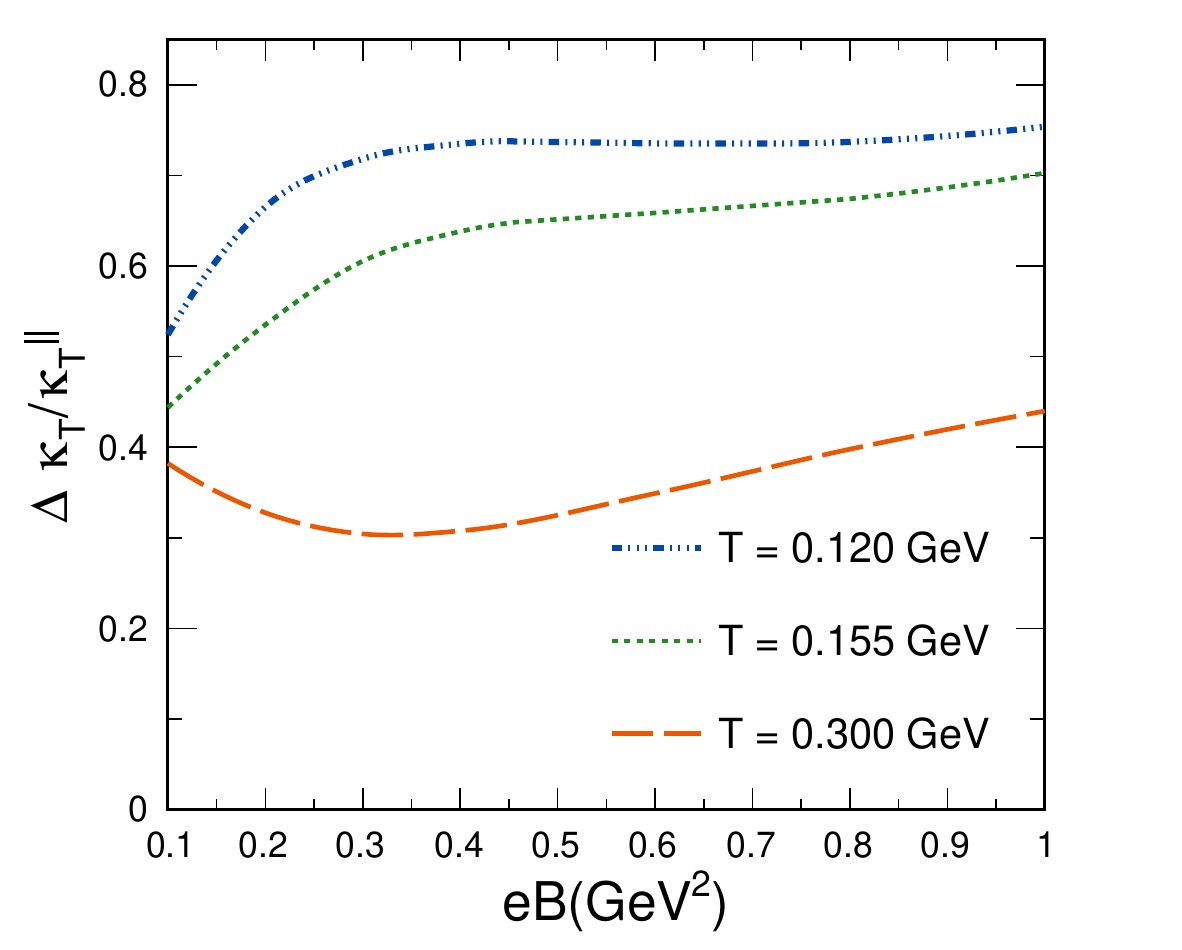}
        \caption{Anisotropy in isothermal compressibility $(\Delta \kappa_{T}/\kappa_T^\parallel)$ as a function of temperature ($T$) (left) and magnetic field ($eB$) (right), $\Delta \kappa_T = \kappa_T^\parallel - \kappa_T^\perp$.}
    \label{IC ratio}
\end{figure*}

The anisotropy between the two components is very prominent. In order to quantify the anisotropy in $\kappa_T$ we have plotted $\Delta \kappa_T/\kappa_T^\parallel$ as a function of $T$ (left) and $eB$ (right) in Fig.~(\ref{IC ratio}), where $\Delta \kappa_T = \kappa_T^\parallel - \kappa_T^\perp$. Similar to $c_s^2$ in Fig.~(\ref{fig speed aniso}), here also we see that anisotropy decreases with increasing $T$ at a fixed $eB$, reaching minima. For temperatures higher than $T_{\rm{c}}$, there is an increase in anisotropic behavior with temperature for each magnetic field value. On the right, we have studied the degree of anisotropy as a function of $eB$. Anisotropy increases with $eB$ at a fixed $T$. Anisotropy is maximum in the strong field zone. Compared to $c_s^2$, anisotropy is higher in $\kappa_T$. In the low $T$ zone, anisotropy in $\kappa_{\rm{T}}$ varies from $60\%$ to $80\%$, and in high $T$, it is $30\%$ to $40\%$ in the mentioned magnetic field regime. In the case of $c_s^2$, the effect is comparatively $10\%$ to $20\%$ less. This signifies that the magnetic field effect is more prominent in the anisotropy of $\kappa_{\rm{T}}$ than that of $c_s^2$.

\section{Summary and outlook}
\label{summary}
In summary, we have estimated the squared speed of sound ($c_s^2$) and isothermal compressibility ($\kappa_T$) of baryon-free QCD matter in the presence of a magnetic field using the PNJL model. With the increasing temperature, $c_s^2$ approaches the conformal limit. It shows a dip around the $T_c$, which shifts to a lower temperature with the increasing magnetic field. The effect of MC and IMC is also found in the components of $c_s^2$ near the transition temperature. The isothermal compressibility is estimated within the PNJL model for the first time. Our results match with other results in the literature qualitatively, confirming that the QCD matter is the most perfect fluid found in nature. With the increasing temperature, $\kappa_T$ decreases. Smooth transition of longitudinal and transverse components from confinement to deconfinement phase indicates crossover transition in the presence of magnetic field at zero chemical potential. Furthermore, we noticed that anisotropy is higher in $\kappa_T$ compared to $c_s^2$, and in the strong field limit, the lowest Landau level is a good approximation. We found that the degree of anisotropy increases with the magnetic field and shows minima near $T_{c}$ when plotted against temperature for both thermodynamical quantities.

The anisotropy in the EoS, due to the magnetic field, is not directly measurable in experiments. However, the observable, directed flow is a promising probe to the initial magnetic field~\cite{STAR:2019clv, ALICE:2019sgg}. Recent studies~\cite{NARA2017543, PhysRevC.106.L061901} show that the directed flow is sensitive to the EoS. The EoS is affected significantly by the external magnetic field. In principle, one should be able to extract information about the effect of the magnetic field on the hot QCD matter by studying the splitting of the directed flow of particle and anti-particle pair. 
Experimentally, at LHC and RHIC, the splitting of the directed flow of $D^{0}$ and $\Bar{D^{0}}$ has been observed~\cite{ALICE:2019sgg, STAR:2019clv}.

As a preliminary study, our work concentrates only on baryon-free QCD matter at a constant magnetic field. This work can be extended to the baryonic matter with a time-dependent magnetic field, which applies to matter created at RHIC and energies achievable below RHIC. Moreover, in this study, the upper limit of the magnetic field is around $1$ GeV owing to the fact that $G(eB)$ is obtained by fitting lQCD data that is available up to $1~\rm{GeV}^{2}$ only. However, the magnetic field generated at RHIC and LHC is around $0.2-0.4 ~\rm{GeV}^{2}$. Thus making $1 ~\rm{GeV}^{2}$ a suitable upper limit. A further theoretical improvement would also enable us to explore the high $eB$ region. In addition, as mentioned in \cite{Fukushima:2003fw}, the dependence of $G$ on $\mu$ and $\Phi$ has been neglected for simplicity. One can find different forms of $G$ in use, but for practical purposes, $G$ should be a function of all the parameters, i.e., $T$, $\mu$, $B$, and $\Phi$, to successfully explain the hot QCD matter. However, considering that the $B$ dependent coupling constant contemplated in this work can reproduce lQCD results qualitatively well, one can proceed with this approximation for further phenomenological studies to explore the QCD medium.

\section{Acknowledgement}
K.G. acknowledges the financial support from the Prime Minister's Research Fellowship (PMRF), Government of India. The authors gratefully acknowledge the DAE-DST, Government of India funding under the mega-science project “Indian participation in the ALICE experiment at CERN” bearing Project No. SR/MF/PS-02/2021-IITI(E-37123).

\bibliographystyle{apsrev4-2}
\bibliography{reference}

\end{document}